# The Characteristic States of the Magnetotelluric Impedance Tensor: Construction, Analytic Properties and Utility in the Analysis of General Earth Conductivity Distributions.

**Andreas Tzanis**

*Department of Geophysics and Geothermy,*
*National and Kapodistrian University of Athens,*
*Panepistimiopoli,*
*Zografou 15784, Greece.*
*E-mail: atzanis@geol.uoa.gr*

Athens, April 2014

**Abstract**

The present work examines basic and interrelated properties of the Magnetotelluric impedance tensor. The first part of the analysis focuses on the local spatial characteristics of impedance tensors obtained at the surface of general Earth conductivity structures. It is demonstrated that the tensor admits an anti-symmetric generalized eigenstate – eigenvalue decomposition consistent with the anti-symmetric interaction between electric and magnetic fields referred to the same coordinate frame. The decomposition is achieved by anti-diagonalization applied through rotation by *two* 2×2 complex operators of the SU(2) rotation group and involves a left operator whose columns comprise the eigenvectors of the electric field and a right operator whose columns comprise the eigenvectors of the total magnetic field; it yields two generalized eigenstates which comprise simple proportional relationships between non-orthogonal, non-horizontal, linearly polarized generalized eigenvalues of the input magnetic and output electric field components (eigen-fields) along the locally fastest (resistive) and slowest (conductive) propagation path into the Earth. These relationships are respectively mediated by the maximum and minimum characteristic values of the impedance tensor (eigen-impedances). The electric and magnetic eigen-fields are three-dimensional and locally orthogonal/ anti-symmetric in complex 3-space; their tilt is respectively a measure of the local landscape of the total electric and magnetic field. When the eigen-fields are projected on the axes of the horizontal observational coordinate frame their components are superimposed and the resulting mixing of phases is manifest in the form of elliptical polarization. The analysis is very closely related to the Singular Value and Canonical Decompositions of the impedance tensor, which are have been constructed on an *ad hoc* basis from a polarization optics point of view and are defined in a different coordinate system.

The analytic properties of the eigen-impedances are next investigated. It is shown from first principles that for linear, source-free Earth structures with arbitrary conductivity distribution but finite conductance, the eigen- impedances are expected to be analytic in the entire lower-half complex frequency plane and their singularities to be confined on the positive imaginary frequency axis so that their real and imaginary parts to be bound by dispersion relations of the Kramers – Kronig kind. In consequence, the eigen-impedances are expected to be positive real (passive) functions and to define freely-decaying induction modes. Insofar as the impedance tensor is produced by isometric transformation of its passive characteristic states, it is also expected to be passive.

The expected passivity of the eigen-impedances is an effective tool by which to appraise measured tensors for compliance with the basic tenets of the Magnetotelluric method. The property can be violated only in the presence of electromagnetic energy sources in the Earth, with dissipation characteristics and time dependence sufficiently different than the respective characteristics of passive induction in a linear medium. This includes all extrinsic effects (e.g. noise). It is also demonstrated with examples, that it includes secondary inductive phenomena generated by realistic conductivity configurations. However, it does *not* include time-independent phenomena taking place in a passive induction context, such as steady-state current channelling, galvanic distortion and electric field twists and reversals. Electric field reversals are given particular attention, as they are frequently blamed for violation of causality. Caution must be exercised in evaluating their effects because the electric field is a *polar* vector with *odd* parity, that is $\mathbf{e}(\mathbf{x}, t) \rightarrow -\mathbf{e}(-\mathbf{x}, t)$; its intrinsic coordinate frame changes relative to the coordinate system of the measurements during a reversal, meaning that the coordinate system in which the impedance tensor is defined also changes. In consequence, a reversal taking place in a passive induction context may be inadvertently interpreted to violate causality and to assert whether a violation has occurred, it is necessary to refer the output electric field to its intrinsic coordinate frame and evaluate the compliance of the eigen-impedances with their expected analytic properties.

The results confirm that passivity is an expected fundamental property of the Magnetotelluric Earth response. The constraints imposed by passivity on the behaviour of the eigen-impedances comprise fundamental tests of the physical validity of measured data and in this respect can be important factors in the process of Magnetotelluric data interpretation.

# 1. Introduction

This presentation comprises two intertwined parts: The first is presented in Section 2 and shows that the (equivalent) *Singular Value Decomposition* (La Torraca et al., 1986) and *Canonical Decomposition* (Yee and Paulson, 1987) of the impedance tensor are, in fact, proper rotations in 3-D space based on the topology of the SU(2) rotation group and produce a characteristic state – characteristic value (generalized eigenstate – eigenvalue) formulation of the Magnetotelluric induction problem. It is also shown that an anti-symmetric reformulation of the decomposition exists, which is suitable for the analysis of Magnetotelluric data when the electric and magnetic fields are referred to the same coordinate frame and is consistent with the conventional rotation applied in typical Magnetotelluric analysis. The second part is presented in Sections 3 and 4 and investigates the basic analytic structure of the characteristic states and values for general Earth conductivity distributions. Specifically, it is shown that in the absence of internal to the Earth sources of electromagnetic energy, the characteristic values (henceforth eigen-impedances) are analytic in the entire lower-half complex frequency plane and their singularities are confined on the positive imaginary frequency axis. It is also demonstrated that these properties can be violated only if there are powerful extrinsic or intrinsic electromagnetic processes with time dependence different than the time dependence of passive Magnetotelluric induction.

The acknowledgment of the exact nature of the SVD and CD is useful because although they offer a unique and powerful analytical and interpretational tool, and have inspired or assisted in the development of other advanced data analysis procedures, they did not have many applications and they are generally not acknowledged for what they actually are and what they do. The main reason may be that they both have been presented as topologically complete *ad hoc* formulations of *polarization states* and not as *intrinsic geometry* of the Magnetotelluric field. Their analysis was (implicitly or explicitly) based on *polarization optics* and their objective was successfully achieved, but many workers on MT may have seen them as only complicated *ad hoc* formulations! Romo et al (2005) verbatim summarize the problem thus: "*… unfortunately, the connection between the resulting earth-responses and the physical model is not as obvious as it is for Swift's analysis of the 2D problem. These difficulties, added to the complexities of the analysis itself, are the probable causes that these methods have not been widely used...*" LaTorraca et al (1986) did not specify that their polarization descriptors are actually SU(2) rotation operators. Yee and Paulson (1987) acknowledge that their decomposition "*… can be viewed as pure or rigid body rotations… [that] can be seen to be the generalization to the complex space of the rotation transformations employed in conventional magnetotelluric analysis…*". However, they leave it at that and do not explain the relationship between the complex and real spaces and the nature of the rotation.

One difficulty in explaining the exact nature of the SVD/CD may be the fact that the theoretical basis (topology and rotation group) is borrowed from quantum mechanics and may thus be regarded "exotic", or even too complex/ inapplicable in the classical context of Magnetotelluric sounding. However, this is not so, if one considers some (usually unappreciated) facts. For instance, the relationship $\mathbf{E}(\omega)=\mathbf{Z}(\omega)\cdot\mathbf{H}(\omega)$, defines the impedance tensor to be a *boson* mediating the exchange of energy between the electric and magnetic fields[1], i.e. the EM interaction in some medium; this boson looks down (into the Earth) and has a spin of 1. Note also that the Magnetotelluric field comprises an ensemble of low frequency – large size photons and $\mathbf{Z}(\omega)$ is the quasi-static analogue of the refractive index. Although we visualize Earth structure in the form of 2-D or 3-D mosaics of macroscopic-to-megascopic size blocks, the much larger photons negotiating their way through the mosaic structure may well think of it as an anisotropic *birefringent* material! In this respect, the rank 2 impedance tensor should embody information about the fast (resistive) and slow (conductive) directions of this material, and should yield it by diagonalization or anti-diagonalization. With such simple arguments, and without going into complex theoretical details, it should be straightforward to see why one may borrow tools from polarization optics and quantum mechanics to analyze the impedance tensor. The SVD/CD formulation *implicitly* adopted the approach of polarization optics (indirect). Given that the analytical tools of polarization optics, especially those describing spatial transformations, are based on the mathematics of spin analysis developed for quantum mechanics, the present analysis attempts to close the circle by taking the second (direct) approach.

The tenet of causality and passivity is intuitive and has *ab initio* been thought to hold for impedance tensors obtained at the surface of any source-free Earth structure. Causality requires that the output of a physical system at time $t$ depends only on the input at times $\tau \leq t$, which implies the vanishing of the output signal at times $t < 0$. Therefore, if the Earth is exclusively excited by external sources $\mathbf{h}_e(t)$, the impedance tensor is causal because $\mathbf{h}_e(t \leq 0) = 0 \Rightarrow \mathbf{h}(t \leq 0) = 0$ and $\mathbf{e}(t \leq 0) = 0$. An input impedance function is further constrained by the requirement of positive energy dissipation: any system that does *not* generate energy, may only dissipate energy when excited by some external input. This implies that the input impedance of that system is bound to be passive and its frequency response to always have a positive real part. It is remarkable nevertheless, that in spite of these simple rules, there have been questions about the causality and passivity of the impedance tensor and for good reasons, as will be discussed herein.

Using a systems theoretical approach and on the premise that Magnetotelluric induction comprises a linear passive system which is automatically *sensu stricto* causal, Yee and Paulson (1988) constructed a Cauer representation of the impedance tensor, which they used to derive twice subtracted dispersion relations (due to the fact that the impedance diverges at high frequencies). Egbert (1990) argued (but did not formally prove) that causality can be violated that these relations lack generality because while for a 1-D Earth the properties of causality and passivity arise naturally from Maxwell's equations, they cannot be valid for general impedance tensors which relate the total (external plus internal) magnetic and output electric fields. He asserted that "*the strong form of passivity assumed by Yee and Paulson almost certainly fails for realistic 3-D conductivity distributions*" and produced contrived examples of conductivity functions that appear to violate causality and passivity, albeit these would generally not be realizable in the Earth's crust (as he also admits to). Egbert also postulated that an indication of causality/passivity violation would be the observation of *anomalous phases*,[2] which he attributed to distortion of the electric field, particularly so when the conductivity structure and/or the distortion produces severe twists or reversals of the orientation of the electric field. He presented examples of anomalous phases, as well as an apparent confirmation of the postulation by forcing the phases of apparently healthy tensors to go out of quadrant after artificially distorting them with real (galvanic) operators. An unfortunate corollary of the breakdown of causality/passivity is that non-passive impedance functions are not interpretable in terms of Earth structure (as will also be discussed in Section 3), ergo impedances exhibiting anomalous phases are not interpretable.

---

[1] This refers to a both a single "interaction" between $\mathbf{E}(\omega)$ and $\mathbf{H}(\omega)$ and the expectation value of multiple interactions; in the latter case, the observed tensor impedance can be thought of as a collection of "bosons" occupying a single state, i.e. a *condensate*!

Anomalous phase observations have been reported by Livelybrooks et al (1996), Pous et al., (2002), Lezaeta and Haak, (2003), Ichihara and Mogi, (2013) and several other authors; they have mostly been explained with models of 3-D induction in near-surface elongate conductors, which may or may not be electromagnetically coupled with other forms of local or regional channelling structures (e.g. faults, ocean, mantle etc.). With a number of such observations, the association of anomalous phases with causality/passivity violations would appear sufficiently verified. However, some recent research appears to reproach this notion. For instance, Ichihara and Mogi (2009) investigated "realistic" L-shaped conductive structures that produce anomalous phases; they attributed the anomalous phases to electric field reversals but concluded that they are not necessarily an effect of distortion. Selway et al. (2012) found a simple explanation for anomalous phases observed in 2-D TE Magnetotelluric data, which indicates that they are only part of the inductive response. In another example, Ichihara et al. (2013) determined that anomalous phases can actually be used to constrain the interpreted resistivity distribution, which means that they are part of the inductive response of the Earth. Anisotropy has also been reported to produce anomalous phases which have almost always been attributed to electric field reversals (for a comprehensive review see Marti, 2014). In a very interesting study, Heise and Pous (2003) produced electric field reversals and anomalous phases with geoelectric models featuring layers of azimuthally anisotropic blocks with different anisotropy strikes embedded in a 2-D regional structure; such Earth conductivity configurations were used for explaining anomalous phases observed in SW Iberia (Pous et al., 2004). In empirical consequence of these investigations, anomalous phases may not necessarily signify the breakdown of causality!

---

[2] In the context of this analysis the phases of measured impedance tensor elements are "*anomalous*" if they transcend the [0°, 90°] or [180°, 270°] intervals in which causality and passivity are supposed to have them confined.

In their rebuttal to Egbert's (1990) comments, Yee and Paulson (1990) argued that even in the case of distortion, passivity and causality are not generally violated and that the subtracted dispersion relations are consistent with Maxwell's equations, unless the system becomes non-linear due to external (source) or internal effects: notwithstanding the possibility of external distortion, the "nature" of the impedance tensor is to be causal when the relationship between electric and magnetic field remains linear. In this author's view, there is a significant difference between Yee and Paulson's (1988, 1990) and Egbert's (1990) and perspectives: The former approached the problem from a systems (filter) theoretical point of view, where the rules governing the behaviour of linear systems are definite and definitive and there can be little doubt what constitutes a causal system and what doesn't. Egbert approached the problem from a physical point of view, trying to show that there are cases where the rules set by filter theory are violated and the property of causality is not general. In order to do so, however, he had to make assumptions about the nature of Earth structure, or about the nature of the processes he considered responsible for the violation of causality (e.g. current channelling).

At any rate, the definition of what is or isn't impedance is quite clear: Impedance is a 'boson' that must effect positive energy dissipation and must therefore be causal and passive. If it doesn't, then it is not impedance and the only reason for this to happen is for sources to exist in the excited system. The rather subtle point here is the definition of "source", which is certainly more general than the concept of a place at which extrinsic electromagnetic energy is produced and injected into the Earth (e.g. anthropogenic current). The very fact that the impedance tensor mediates a passive process clearly indicates that "source" can be *any* process with time dependence different than the time dependence of passive induction and amplitude sufficiently high as to disturb it and make its presence known. This *does not* rule out secondary inductive effects generated within the Earth but could rule out processes without time dependence, such as galvanic distortion!

In order to be more specific, consider that the system of electromagnetic interactions within the Earth comprises passive induction and "other" external or internal effects like channelling, distortion etc. All these take place in a linear conductivity medium, meaning that they are additive (superimposable) *regardless* of causality. For an $\exp(+i\omega t)$ harmonic dependence, a transfer function is *sensu stricto* causal if all the singularities of its frequency transform are exclusively located in the upper (positive imaginary) half of the complex frequency plane or, equivalently, if all the singularities of the input and output processes are located in the upper (positive imaginary) half of the complex frequency plane. A combination of causal and non-causal linear processes is governed by the parallel filter rule, (e.g. Claerbout, 1976; Robinson 1980), which states that the frequency transform of the combined system has the same number of singularities in the upper-half frequency plane, as the transform of the dominant process. If the combining processes are strictly causal, then the combination will be strictly causal. If it is not, the combination will develop zeros in the lower-half frequency plane and depending on their number, may wax and wane between *sensu lato* causality and complete instability. Exactly the opposite holds for an exp($-i\omega t$) harmonic dependence. In other words, non-causal and causal processes may co-exist in a linear system and the result may be causal provided the magnitude and phase spectrum of the non-causal process is less than that of the causal processes at all frequencies. Claerbout has summarized this property in his "garbage theorem" thus: "*You can add garbage to a minimum phase wavelet if you do not add too much*" (Claerbout, 1976, p31).

According to the parallel filter rule, causality and passivity will break down only if there are dominant non-causal processes operating in the Earth system. Effects such as electric field reversals caused by internal processes are not at all certain to violate passivity. The point here is that while the magnetic field is an *axial* vector with even parity under spatial inversions, behaving like **h**(**x**, *t*) → **h**′(**x**, *t*) = +**h**(−**x**, *t*), the electric field is a *polar* vector with *odd* parity so that **e**(**x**, *t*) → **e**′(**x**, *t*) = −**e**(−**x**, *t*), e.g. Jackson (1975, pp247-249). The complication is that the intrinsic coordinate system of the electric field changes relative to the coordinate system of the measurements! Thus, a reversal in the orientation of the electric field may only result in *apparent* violation of passivity because the tensor is now defined in a different coordinate system than that of the measurements! If the reversal is part of the passive induction process it may lead to a shift of the electric field's phase between stable states but to no change of the time dependence, i.e. no output at *t* < 0. If it is associated with violation of passivity, then the time-dependence of its source must be both non-passive and dominant and must remain so during the transition from the normal to the reversed state.

The above discussion indicates that the question of causality and passivity violation of the impedance tensor (and its association with anomalous phase observations) is still open and in need of clarification with systematic examination from first principles. An attempt to this end is presented herein. Because the characteristic state – characteristic value formulation constructed in Section 2, offers a *compact* and unique means to characterize the impedance tensor, the analysis of Sections 3 and 4 will examine their utility in establishing objective conditions and measures of the tensor's causality and passivity, without assumptions about the conductivity structure, but only on the basis of (pre)Maxwell equations, absence of internal to the Earth sources of electromagnetic energy and uniform external source fields. An answer to this question would determine under which conditions causality may be violated and would formulate a first principles approach for testing the compliance of measured data with the basic tenets of the Magnetotelluric method.

# 2. The nature of 3-D Impedance Tensor decompositions.

## 2.1 Definitions and Rotation Operators

The basic tool to be used in the ensuing analysis is the SU(2) rotation group and its irreducible representations. The subject is very involved and extensive; for a very good introduction to the group and its applications, the reader can see Arfken and Weber (2005). Advanced and detailed studies can be found in Murnaghan (1938) and Normand (1980), while Wigner (1959) and Rose (1957) present readable accounts of the rotation group. Only elementary and absolutely essential information is given here.

The group SU(2) is a continuous, compact subset of the U($n$) Lie group of $n \times n$ unitary matrices with $n^2-1$ independent parameters. The condition $\det\{U(n)\}=+1$ imposes rotations only, no reflections and defines the Special Unitary (Unimodular) group SU($n$). For $n=2$ there exist three independent parameters (that amount to rotation angles). Whereas (*) denotes complex conjugation and (†) Hermitian transposition, the rotation operators of SU(2) are $2\times 2$ complex matrices of the form

$$\mathbf{U} = \begin{bmatrix} \alpha & \beta \\ -\beta^* & \alpha^* \end{bmatrix},$$

so that $\det\{\mathbf{U}\}=|\alpha|^2+|\beta|^2=1$ and $\mathbf{U}^\dagger = \mathbf{U}^{-1}$ $\forall$ $\mathbf{U} \in$ SU(2).

The vector space structure is Euclidean. Consider the three anti-commuting Pauli spin matrices

$$\mathbf{s}_1 = \begin{bmatrix} 0 & 1 \\ 1 & 0 \end{bmatrix}, \quad \mathbf{s}_2 = \begin{bmatrix} 0 & i \\ -i & 0 \end{bmatrix}, \quad \mathbf{s}_3 = \begin{bmatrix} 1 & 0 \\ 0 & -1 \end{bmatrix} \tag{1}$$

with $(\mathbf{s}_j)^2=\mathbf{I}$, obeying the commutation relationships $-i[\mathbf{s}_i,\mathbf{s}_j]=2\varepsilon_{ijk}\mathbf{s}_k$ where $[\mathbf{a},\mathbf{b}]=\mathbf{ab}-\mathbf{ba}$. These matrices are Hermitian, traceless and linearly independent over the real field; together with $\mathbf{I}$ they form a complete set and due to their commutation relationships they form an orthonormal vector basis of the 3-D space, over the real field.

The familiar three-dimensional space (3-space) we live in is defined over the real field $\mathbb{R}^3$. Rotations in $\mathbb{R}^3$ are specified by representations of the Special Orthogonal Lie group SO(3) of $3 \times 3$ real valued unimodular matrices. It can be shown that from any Cartesian tensor in $\mathbb{R}^3$, one can define a mapping onto the set of 2x2 complex matrices, in the Hilbert space of complex valued $L^2$ (squared) functions on $\mathbb{R}^3$, which for the purpose of spin (rotation) analysis *only*, reduces to $\mathbb{C}^2$; this we can visualize in the simple zero-trace Hermitian form

$$\mathbf{P}(x,y,z) = \mathbf{s}_1 x + \mathbf{s}_2 y + \mathbf{s}_3 z = \begin{vmatrix} \mathbf{z} & \mathbf{x}+i\mathbf{y} \\ \mathbf{x}-i\mathbf{y} & -\mathbf{z} \end{vmatrix} \tag{2}$$

with $\det\{\mathbf{P}\}=x^2+y^2+z^2=1$. SU(2) enters as a symmetry group in $\mathbb{C}^2$. For any unimodular matrix $\mathbf{U} \in$ SU(2), an arbitrary unitary transformation $\mathbf{P} \to \mathbf{Q} = \mathbf{U}\cdot\mathbf{P}\cdot\mathbf{U}^\dagger$ is also traceless Hermitian, so that $\mathbf{Q}(x', y', z') = \mathbf{s}_1 x' + \mathbf{s}_2 y' + \mathbf{s}_3 z'$. Since $\det\mathbf{Q} = \det\mathbf{P}$, the real linear transformation $\{x,y,z\} \to \{x', y', z'\}$ induced by $\mathbf{P} \to \mathbf{Q} = \mathbf{P}(x', y', z')$ is such that $\mathbf{x}^2 + \mathbf{y}^2 + \mathbf{z}^2 = \mathbf{x'}^2 + \mathbf{y'}^2 + \mathbf{z'}^2$. In other words,

$$\begin{bmatrix} x' \\ y' \\ z' \end{bmatrix} = \mathbf{O} \cdot \begin{bmatrix} x \\ y \\ z \end{bmatrix},$$

where $\mathbf{O}$ is a *real* orthogonal 3x3 matrix comprising a representation of group SO(3). It is easy to show that any subsequent unitary transformation $\mathbf{Q} \to \mathbf{R} = \mathbf{V} \cdot \mathbf{Q} \cdot \mathbf{V}^\dagger$ induces the orthogonal transformation

$$\begin{bmatrix} x' \\ y' \\ z' \end{bmatrix} \to \begin{bmatrix} x'' \\ y'' \\ z'' \end{bmatrix} = \mathbf{O}(\mathbf{V}) \cdot \begin{bmatrix} x' \\ y' \\ z' \end{bmatrix} = \mathbf{O}(\mathbf{V}) \cdot \mathbf{O}(\mathbf{U}) \begin{bmatrix} x \\ y \\ z \end{bmatrix},$$

or, in other words, $\mathbf{O}(\mathbf{VU}) = \mathbf{O}(\mathbf{V})\, \mathbf{O}(\mathbf{U})$.

This demonstrates in a naïve albeit instructive manner, that the collection of 3x3 real orthogonal matrices $\mathbf{O}(\mathbf{U})$ obtained by letting $\mathbf{U}$ wander over the 2x2 unitary group SU(2) constitutes a representation of the 3x3 rotation group SO(3) by 2x2 unitary matrices. Note though that SO(3) and SU(2) are *only locally isomorphic*. This means that as long as small rotations are considered, one is not able to tell the difference. However, a rotation of 360° corresponds to an element of SU(2) that is not identity. Thus, the representation is *homeomorphic*, in the following sense: $\{\mathbf{O}(\mathbf{U})\}$ is a representation of SU(2) having the property $\mathbf{O}(\mathbf{U}) = \mathbf{O}(\mathbf{V}) \Leftrightarrow \mathbf{V} = \pm\mathbf{U}$, i.e. the rotations will be unique to within a symmetry of $2\pi$. Thus, SU(2) emerges as the *universal covering space* of SO(3), with covering map 2 → 1 (a double cover) and the topology of the 3-sphere $S^3$ (i.e. four dimensional).

The SU(2) group elements can be generated from the Pauli matrices by expanding the exponential in a Maclaurin series and separating even and odd powers, that is

$$\mathbf{U}_j = \exp\left(\pm i \frac{\psi}{2} \mathbf{s}_j\right) \approx \cos\left(\frac{\psi}{2}\right) \mathbf{I} \pm i \sin\left(\frac{\psi}{2}\right) \mathbf{s}_j, \tag{3}$$

where $\psi$ is a real parameter (rotation angle). However, the choice of the rotation operators for the analysis of the tensor impedance requires some caution: it has to be *natural* with respect to the Magnetotelluric induction problem otherwise the numerical estimation procedures will not work.

The electric and magnetic coordinate frames commonly used for Magnetotelluric data acquisition are right-handed with the *y*-axis positive to the right of the *x*-axis and the *z*-axis pointing into the Earth; a natural choice is the 'geomagnetic' reference frame with *x*-North, *y*-East, *z*-Down. In such frames, the positive sign in the exponential of equation (3) will correspond to clockwise rotations. It is significant to point out that this coordinate system is different from the standard used in polarization optics (*x*-right, *y*-top and *z*-up), meaning that due adjustments should be made if comparisons are to be attempted (for instance, the rotation operators should be generated by choosing the minus sign in eq. (3), etc.).

Given the above definitions, a rotation about the *z*-axis is performed by the operator:

$$\mathbf{U}_z = \exp\left(i\frac{\varphi}{2}\mathbf{s}_3\right) \approx \begin{bmatrix} e^{i\frac{\varphi}{2}} & 0 \\ 0 & e^{-i\frac{\varphi}{2}} \end{bmatrix}.$$

However, an easy calculation yields

$$\mathbf{U}_z \mathbf{s}_1 \mathbf{U}_z^\dagger = \begin{bmatrix} 0 & e^{i\varphi} \\ e^{-i\varphi} & 0 \end{bmatrix} = \cos\varphi \cdot \mathbf{s}_1 + \sin\varphi \cdot \mathbf{s}_2 \quad \Rightarrow \mathbf{x}' = \cos\varphi \cdot \mathbf{x} - \sin\varphi \cdot \mathbf{y}$$

$$\mathbf{U}_z \mathbf{s}_2 \mathbf{U}_z^\dagger = \begin{bmatrix} 0 & ie^{i\varphi} \\ -ie^{-i\varphi} & 0 \end{bmatrix} = -\sin\varphi \cdot \mathbf{s}_1 + \cos\varphi \cdot \mathbf{s}_2 \quad \Rightarrow \mathbf{y}' = \sin\varphi \cdot \mathbf{x} + \cos\varphi \cdot \mathbf{y}$$

$$\mathbf{U}_z \mathbf{s}_3 \mathbf{U}_z^\dagger = \begin{bmatrix} 1 & 0 \\ 0 & -1 \end{bmatrix} = \mathbf{s}_3 \quad \Rightarrow \mathbf{z}' = \mathbf{z}.$$

Thus, in the basis **x** + *i***y** ,

$$\mathbf{U}_z = \exp\left(i\frac{\varphi}{2}\mathbf{s}_3\right) \approx \begin{bmatrix} \cos\varphi & -\sin\varphi \\ \sin\varphi & \cos\varphi \end{bmatrix} \in \mathrm{SO}(2) \subset \mathrm{SU}(2). \tag{4}$$

Eq. (4) is the familiar rotation operator introduced by Word et al (1970) and customarily used in conventional Magnetotelluric analysis. It is well known that this operator cannot reduce a 3-D tensor impedance to its principal components, i.e. to a pure diagonal or anti-diagonal form, in the space **P**(*x, y, z*). Using foresight (and pending justification) an additional rotation is required about one of the horizontal axes. The obvious choice is the (major) *x′*-axis of the fast direction, relative to which the minor (slow) axis is defined. To this effect, a clockwise rotation about the *x*-axis is performed with the operator

$$\mathbf{U}_x = \exp\left(i\frac{\vartheta}{2}\mathbf{s}_1\right) \approx \begin{bmatrix} \cos\frac{\vartheta}{2} & i\sin\frac{\vartheta}{2} \\ i\sin\frac{\vartheta}{2} & \cos\frac{\vartheta}{2} \end{bmatrix},$$

whence

$$\mathbf{U}_x \mathbf{s}_1 \mathbf{U}_x^\dagger = \begin{bmatrix} 0 & 1 \\ 1 & 0 \end{bmatrix} = \mathbf{s}_1 \quad \Rightarrow \mathbf{x}' = \mathbf{x}$$

$$\mathbf{U}_x \mathbf{s}_2 \mathbf{U}_x^\dagger = \begin{bmatrix} \sin\vartheta & i\cos\vartheta \\ -i\cos\vartheta & -\sin\vartheta \end{bmatrix} = \cos\vartheta \cdot \mathbf{s}_2 + \sin\vartheta \cdot \mathbf{s}_3 \quad \Rightarrow \mathbf{y}' = \cos\vartheta \cdot \mathbf{y} - \sin\vartheta \cdot \mathbf{z}$$

$$\mathbf{U}_x \mathbf{s}_3 \mathbf{U}_x^\dagger = \begin{bmatrix} \cos\vartheta & -i\sin\vartheta \\ i\sin\vartheta & -\cos\vartheta \end{bmatrix} = -\sin\vartheta \cdot \mathbf{s}_2 + \cos\vartheta \cdot \mathbf{s}_3 \quad \Rightarrow \mathbf{z}' = \sin\vartheta \cdot \mathbf{y} + \cos\vartheta \cdot \mathbf{z}$$

Thus, letting $\theta = \vartheta/2$, a clockwise rotation about the z-axis, followed by a clockwise rotation about the x-axis is

$$\mathbf{U}_{zx}(\varphi,\vartheta) = \mathbf{U}_z(\varphi)\mathbf{U}_x(\vartheta) = \begin{bmatrix} \cos\varphi & -\sin\varphi \\ \sin\varphi & \cos\varphi \end{bmatrix} \cdot \begin{bmatrix} \cos\theta & i\sin\theta \\ i\sin\theta & \cos\theta \end{bmatrix}. \tag{5}$$

In the foregoing presentation, only rotations of the coordinate system where assumed. However, a very important property of SU(2), is that it comprises operators capable of rotating complex functions $f(x, y, z)$ of a fixed coordinate system. By definition, the operation $\mathbf{U}\{f(x, y, z)\}$ creates a new function $f'(x, y, z)$, which is numerically equal to $f(x', y', z')$, where the primes indicate that the coordinate system has been rotated by $\mathbf{U}$. One can show that $\mathbf{U}_j(\psi) = \exp(i\psi L_j)$ where $L_j$ are the cartesian components

$$iL_1 = \left\{ \mathbf{y}\frac{\partial}{\partial x} - \mathbf{z}\frac{\partial}{\partial y} \right\}, \quad iL_2 = \left\{ \mathbf{z}\frac{\partial}{\partial x} - \mathbf{x}\frac{\partial}{\partial z} \right\}, \quad iL_3 = \left\{ \mathbf{x}\frac{\partial}{\partial y} - \mathbf{y}\frac{\partial}{\partial z} \right\}$$

of the quantum mechanical angular momentum operator. The components $L_j$ obey the commutation relationship $[L_i, L_j] = i\varepsilon_{ijk}L_k$ or $\mathbf{L}\times\mathbf{L} = i\mathbf{L}$, which are similar to those for the SU(2) vector basis. The similarity is not coincidental; it can be shown that $\mathbf{U}_j = \exp(i\frac{1}{2}\psi\mathbf{s}_j)$ are the rotation operators generated by the angular momentum operator $\mathbf{L}_j$ (e.g. Rose, 1957, pp 22-29).

## 2.2. Rotation and Decomposition

This presentation is concerned with the (anti)diagonalization of the impedance tensor via rotation (coordinate system transformation). It is therefore imperative to clarify the geometry of the coordinate system(s) in which the impedance tensor is defined and which will be transformed. In general, owing to the orthogonality of the electric and magnetic fields (and electromagnetic induction thereof), the impedance tensor may be defined in one of two *right-handed* coordinate systems:

- In the *Coordinate System 1* (CS-1), the horizontal axes of the magnetic (input) coordinate frame $(x_h, y_h)$ are rotated by 90° clockwise with respect to the horizontal axes of the electric (output) reference frame $(x_e, y_e)$ according to

$$\begin{bmatrix} x_h \\ y_h \end{bmatrix} = \begin{bmatrix} 0 & -1 \\ 1 & 0 \end{bmatrix} \begin{bmatrix} x_e \\ y_e \end{bmatrix} = \mathbf{R}\left(\frac{\pi}{2}\right)\begin{bmatrix} x_e \\ y_e \end{bmatrix},$$

so that the $x_h$-axis is parallel to the $y_e$-axis and the $y_h$-axis anti-parallel to the $x_e$-axis. In this system, the relationship (mapping) between the transverse components of the magnetic input and electric output fields is

$$\begin{bmatrix} E_{x_e}(\omega) \\ E_{y_e}(\omega) \end{bmatrix} = \begin{bmatrix} \mathcal{Z}_{x_e x_h}(\omega) & \mathcal{Z}_{x_e y_h}(\omega) \\ \mathcal{Z}_{y_e x_h}(\omega) & \mathcal{Z}_{y_e y_h}(\omega) \end{bmatrix}\begin{bmatrix} H_{x_h}(\omega) \\ H_{y_h}(\omega) \end{bmatrix} \quad \leftrightarrow \quad \mathbf{E}(\omega) = \mathcal{Z}(\omega)\underline{\mathbf{H}}(\omega) \tag{6}$$

and is apparently *symmetric*. CS-1 is seldom (if at all) used in Magnetotelluric practice but has been implemented in fundamental theoretical work because the symmetric input – output mapping facili-tates the direct application of physical and mathematical concepts known from the analysis of symmetric physical systems, to the Magnetotelluric problem: it is the coordinate system used by Yee and Paulson (1987) and implicitly by LaTorraca et al (1986), although the latter did not specify this detail.

- In the *Coordinate System 2* (CS-2), the input magnetic and output electric frames are identical, that is $(x_h, y_h) \equiv (x_e, y_e) \equiv (x, y)$, and the transverse magnetic input and electric output field components are associated with the familiar relationship

$$\begin{bmatrix} E_x(\omega) \\ E_y(\omega) \end{bmatrix} = \begin{bmatrix} Z_{xx}(\omega) & Z_{xy}(\omega) \\ Z_{yx}(\omega) & Z_{yy}(\omega) \end{bmatrix} \begin{bmatrix} H_x(\omega) \\ H_y(\omega) \end{bmatrix} \leftrightarrow \mathbf{E}(\omega) = \mathbf{Z}(\omega)\mathbf{H}(\omega),$$

which is apparently *anti-symmetric*. CS-2 is the system commonly used in Magnetotelluric practice. The impedance tensors $\mathcal{Z}(\omega)$ and $\mathbf{Z}(\omega)$ are related as

$$\mathcal{Z}(\omega) = \mathbf{Z}(\omega)\mathbf{R}\left(\tfrac{\pi}{2}\right) \tag{7}$$

Now, consider that in either system, a rotation by a single operator $\mathbf{U}(\theta, \varphi) \equiv \mathbf{U}_{zx}(\theta, \varphi)$, of the form $\mathbf{U}^\dagger(\theta,\varphi) \cdot \mathbf{X} \cdot \mathbf{U}(\theta,\varphi)$ *cannot* reduce $\mathcal{Z}(\omega)$ or $\mathbf{Z}(\omega)$ to diagonal or anti-diagonal forms. The necessary and sufficient condition for a complex matrix $\mathbf{X}$ to be diagonalizable by a single unitary operator is to be normal so that $[\mathbf{X}, \mathbf{X}^\dagger] = 0$. In the general case the impedance tensor is *regular*, not normal (symmetric) and depends on eight independent parameters (degrees of freedom or topological dimensions), where each of $\Re\{Z_{ij}\}$ and $\Im\{Z_{ij}\}$ is assigned with one degree of freedom. Therefore, assuming that (anti)diagonalization could be done with such an operation, it would depend on a maximum of six independent parameters out of the eight existing in the tensor, i.e. four in the two complex principal impedances plus two rotation angles. Therefore, it would be *incomplete* – 'topologically deficient' so to speak. It follows that exactly two operators $\mathbf{U}(\theta_1, \varphi_1)$ and $\mathbf{V}(\theta_2, \varphi_2)$ are required to (anti)diagonalize the impedance tensor, thereby providing an eight parameter set that completely describes it (four in the two complex principal impedance and four rotation angles). To spare space, time and argument, it is possible to justify the application of the two-operator rotation on the basis of the following Propositions:

**Proposition 1:** If $\mathbf{X}$ is a regular $n \times n$ complex matrix of rank $n$, then there exist one $n \times n$ unitary matrix $\mathbf{U}$, one $n \times n$ unitary matrix $\mathbf{V}$ and one $n \times n$ *diagonal* complex matrix $\mathbf{S}$ such that $\mathbf{X} = \mathbf{U}\mathbf{S}\mathbf{V}^\dagger$. The successive diagonal elements of $\mathbf{S}$ are *non-increasing*.

*Proof:* The matrix $\mathbf{X}\mathbf{X}^\dagger$ is normal and has an eigenvalue – eigenvector decomposition $\mathbf{X}^\dagger\mathbf{X} = \mathbf{V}\mathbf{D}\mathbf{V}^\dagger$ where $\mathbf{V}$ is $n \times n$ unitary and $\mathbf{D}$ is diagonal with positive entries. The matrix $\mathbf{D}$ is defined to be the product $\mathbf{S}^\dagger\mathbf{S}$ with $D_{ii} > D_{jj}$ for $i > j$. Let $\mathbf{Y} = \mathbf{S}^{-1}$. Then, it can be easily be verified that $\mathbf{Y}^\dagger\mathbf{D}\mathbf{Y} = \mathbf{I}_n$. Define the $n \times n$ matrix $\mathbf{U} = \mathbf{X}\mathbf{V}\mathbf{Y}$ so that $\mathbf{U}^\dagger\mathbf{U} = \mathbf{Y}^\dagger\mathbf{V}^\dagger\mathbf{X}^\dagger\mathbf{X}\mathbf{V}\mathbf{Y} = \mathbf{Y}^\dagger\mathbf{D}\mathbf{Y} = \mathbf{I}_n$, which shows that $\mathbf{U}$ is unitary. Then again, from the definitions $\mathbf{U}\mathbf{S}\mathbf{V}^\dagger = \mathbf{X}\mathbf{V}\mathbf{Y}\mathbf{S}\mathbf{V}^\dagger = \mathbf{X}\mathbf{V}\mathbf{V}^\dagger = \mathbf{X}$ because $\mathbf{V}$ is unitary. Thus, it is shown that the decomposition is realizable.

This type of diagonalization possesses all the properties of the Singular Value Decomposition extended to regular complex square matrices and unitary operators. It should not be news to anyone because its existence has already been documented in Magnetotelluric literature, (e.g. LaTorraca et al, 1986), although it was not formally proved. The proof was given here for the sake of completeness.

**Proposition 2:** If **X** is a regular $n{\times}n$ complex matrix of rank $n$, then there exist one $n{\times}n$ unitary matrix **U**, one $n{\times}n$ unitary matrix **V** and one $n{\times}n$ *anti-diagonal* complex matrix **S** such that $\mathbf{X} = \mathbf{U}\mathbf{S}\mathbf{V}^{\dagger}$. The successive anti-diagonal elements of **S** are *non-decreasing*.

*Proof:* The matrix $\mathbf{X}\mathbf{X}^{\dagger}$ is normal and has an eigenvalue – eigenvector decomposition $\mathbf{X}\mathbf{X}^{\dagger} = \mathbf{U}\mathbf{D}\mathbf{U}^{\dagger}$ where **U** is $n{\times}n$ unitary and **D** is diagonal with positive entries. The matrix **D** is defined to be the product $\mathbf{S}\mathbf{S}^{\dagger}$ with $D_{ii} > D_{jj}$ for $i > j$. Then, it can be easily be verified that $\mathbf{S}^{-1}\mathbf{D}(\mathbf{S}^{-1})^{\dagger} = \mathbf{I}_n$. Now define the $n{\times}n$ matrix $\mathbf{V}^{\dagger} = \mathbf{S}^{-1}\mathbf{U}^{\dagger}\mathbf{X}$, so that $\mathbf{V} = \mathbf{X}^{\dagger}\mathbf{U}(\mathbf{S}^{-1})^{\dagger}$ and $\mathbf{V}^{\dagger}\mathbf{V} = \mathbf{S}^{-1}\mathbf{U}^{\dagger}\mathbf{X}\mathbf{X}^{\dagger}\mathbf{U}(\mathbf{S}^{-1})^{\dagger} = \mathbf{S}^{-1}\mathbf{D}(\mathbf{S}^{-1})^{\dagger} = \mathbf{I}_n$, which shows that **V** is unitary. Then again from the definitions, $\mathbf{U}\mathbf{S}\mathbf{V}^{\dagger} = \mathbf{U}\mathbf{S}\mathbf{S}^{-1}\mathbf{U}^{\dagger}\mathbf{X} = \mathbf{U}\mathbf{U}^{\dagger}\mathbf{X} = \mathbf{X}$ because **U** is unitary. It is thus shown that the decomposition is realizable.

Now, owing to the fact that **S** is anti-diagonal, $\mathbf{S}\mathbf{S}^{\dagger} \neq \mathbf{S}^{\dagger}\mathbf{S}$ and it is necessary to verify that in order to be unique, the same decomposition is realizable for the product $\mathbf{X}^{\dagger}\mathbf{X}$. Again, let the matrix $\mathbf{X}^{\dagger}\mathbf{X}$ admit the eigenvalue – eigenvector decomposition $\mathbf{X}^{\dagger}\mathbf{X} = \mathbf{V}\mathbf{G}\mathbf{V}^{\dagger}$ with **V** unitary. Let $\mathbf{G} = \mathbf{S}^{\dagger}\mathbf{S}$ with $G_{ii} < G_{jj}$ for $i > j$, whence it is easily verified that $(\mathbf{S}^{-1})^{\dagger}\mathbf{G}\mathbf{S}^{-1} = \mathbf{I}_n$. Now, from the definition of **V** given above, it follows that $\mathbf{U}^{\dagger} = \mathbf{S}\mathbf{V}^{\dagger}\mathbf{X}^{-1}$ and $\mathbf{U} = (\mathbf{X}^{-1})^{\dagger}\mathbf{V}\mathbf{S}^{\dagger}$. Then, $\mathbf{U}\mathbf{U}^{\dagger} = (\mathbf{X}^{-1})^{\dagger}\mathbf{V}\mathbf{S}^{\dagger}\mathbf{S}\mathbf{V}^{\dagger}\mathbf{X}^{-1} = (\mathbf{X}^{-1})^{\dagger}\mathbf{V}\mathbf{V}^{\dagger}\mathbf{X}^{\dagger}\mathbf{X}\mathbf{V}\mathbf{V}^{\dagger}\mathbf{X}^{-1} = (\mathbf{X}^{-1})^{\dagger}\mathbf{X}^{\dagger}\mathbf{X}\mathbf{X}^{-1}$ and because $(\mathbf{X}^{-1})^{\dagger} = (\mathbf{X}^{\dagger})^{-1}$ one obtains $\mathbf{U}\mathbf{U}^{\dagger} = \mathbf{I}_n$. This shows that **U** is unitary and unique. Finally, from the definition $\mathbf{U}\mathbf{S}\mathbf{V}^{\dagger} = (\mathbf{X}^{-1})^{\dagger}\mathbf{V}\mathbf{S}^{\dagger}\mathbf{S}\mathbf{V}^{\dagger} = (\mathbf{X}^{-1})^{\dagger}\mathbf{X}^{\dagger}\mathbf{X} = \mathbf{X}$, which also shows that the decomposition is unique and realizable for the unitary matrices **U** and **V**.

Let us first construct the diagonal (symmetric) rotation and decomposition of the tensor $\mathcal{Z}(\omega)$ allowed by Proposition 1. The products $\mathcal{C}_1(\omega) = \mathcal{Z}(\omega)\mathcal{Z}^{\dagger}(\omega)$ and $\mathcal{C}_2(\omega) = \mathcal{Z}^{\dagger}(\omega)\mathcal{Z}(\omega)$ are normal (Hermitian) matrices and constitute mappings of $\mathcal{Z}(\omega)$ onto $\mathbb{C}^2$. Their norms are equal, but $\left[\mathcal{C}_1(\omega), \mathcal{C}_2^{\dagger}(\omega)\right] \neq \mathbf{0}$ while $\left[\mathcal{C}_j(\omega), \mathcal{C}_j^{\dagger}(\omega)\right] = \mathbf{0}$; thus $\mathcal{C}_1(\omega)$ and $\mathcal{C}_2(\omega)$ encode different pieces of the geometrical information originally stored in $\mathcal{Z}(\omega)$ and as will become clear later on, this pertains to the characteristic coordinate frames of the electric and magnetic field respectively. Moreover, each of $\mathcal{C}_j$ depends on only four degrees of freedom, meaning that each can be diagonalized with a single unitary rotation operator of the form (5). Accordingly, $\mathcal{C}_1(\omega)$ and $\mathcal{C}_2(\omega)$ admit *eigenvalue-eigenvector* decompositions of the form

$$\mathcal{C}_1(\omega) = \mathbf{U}(\theta_1, \varphi_1, \omega) \cdot \mathcal{D}(\omega) \cdot \mathbf{U}^\dagger(\theta_1, \varphi_1, \omega) = \mathbf{U}(\theta_1, \varphi_1, \omega) \cdot \begin{bmatrix} r_1^2(\omega) & 0 \\ 0 & r_2^2(\omega) \end{bmatrix} \cdot \mathbf{U}^\dagger(\theta_1, \varphi_1, \omega) \quad (8a)$$

and

$$\mathcal{C}_2(\omega) = \mathbf{V}(\theta_2, \varphi_2, \omega) \cdot \mathcal{D}(\omega) \cdot \mathbf{V}^\dagger(\theta_2, \varphi_2, \omega) = \mathbf{V}(\theta_2, \varphi_2, \omega) \cdot \begin{bmatrix} r_1^2(\omega) & 0 \\ 0 & r_2^2(\omega) \end{bmatrix} \cdot \mathbf{V}^\dagger(\theta_2, \varphi_2, \omega) \quad (8b)$$

Now define the complex diagonal tensor

$$\tilde{\mathcal{Z}}(\omega) = \begin{bmatrix} \zeta_1(\omega) & 0 \\ 0 & \zeta_2(\omega) \end{bmatrix}, \quad |\zeta_1(\omega)| > |\zeta_2(\omega)|, \quad \zeta_j(\omega)\zeta_j^*(\omega) = r_j^2, \quad j = 1, 2$$

to be the *characteristic impedance* or, more precisely, the *eigen-impedance* tensor, so that

$$\tilde{\mathcal{Z}}(\omega)\tilde{\mathcal{Z}}^\dagger(\omega) = \tilde{\mathcal{Z}}^\dagger(\omega)\tilde{\mathcal{Z}}(\omega) = \mathcal{D}(\omega)$$

Then, by virtue of Proposition 1,

$$\tilde{\mathcal{Z}}(\omega) = \mathbf{U}^\dagger(\theta_1, \varphi_1, \omega) \cdot \mathcal{Z}(\omega) \cdot \mathbf{V}(\theta_2, \varphi_2, \omega) \quad (8c)$$

whence it is straightforward to obtain (8a) and (8b) by direct multiplication. Eq. (8c) is precisely the Singular Value Decomposition of LaTorraca et al. (1986) *and* the Canonical Decomposition[3] of Yee and Paulson (1987) respectively.

Now, right multiply eq. (8c) by $\mathbf{R}\left(-\frac{\pi}{2}\right) = \begin{bmatrix} 0 & 1 \\ -1 & 0 \end{bmatrix}$ to rotate the eigen-impedance tensor from CS-1 to CS-2 and on substituting eq. (7) obtain

$$\tilde{\mathcal{Z}}(\omega)\mathbf{R}\left(-\tfrac{\pi}{2}\right) = \mathbf{U}^\dagger(\theta_1, \varphi_1, \omega)\mathbf{Z}(\omega)\mathbf{R}\left(\tfrac{\pi}{2}\right)\mathbf{V}(\theta_2, \varphi_2, \omega)\mathbf{R}\left(-\tfrac{\pi}{2}\right), \quad (9)$$

with

$$\tilde{\mathbf{Z}}(\omega) = \tilde{\mathcal{Z}}(\omega)\mathbf{R}\left(-\tfrac{\pi}{2}\right) = \begin{bmatrix} 0 & \zeta_1(\omega) \\ -\zeta_2(\omega) & 0 \end{bmatrix}$$

comprising the eigen-impedance tensor in CS-2. Moreover,

$$\mathbf{R}\left(\tfrac{\pi}{2}\right)\mathbf{V}(\theta_2, \varphi_2, \omega)\mathbf{R}\left(-\tfrac{\pi}{2}\right) = \mathbf{R}\left(\tfrac{\pi}{2}\right)\mathbf{V}_z(\varphi_2, \omega)\mathbf{V}_x(\theta_2, \omega)\mathbf{R}\left(-\tfrac{\pi}{2}\right)$$

---

[3] More precisely the reformulated decomposition specified in page 179. The basic form shown in page 177 cannot be realized numerically.

Because $\mathbf{R}\left(\pm\frac{\pi}{2}\right) \in \mathrm{SO}(2) \subseteq \mathrm{SU}(2)$ and $\mathbf{V}_z \in \mathrm{SO}(2) \subseteq \mathrm{SU}(2)$, we have $\left[\mathbf{R}\left(\pm\frac{\pi}{2}\right), \mathbf{V}_z\right] = 0$: the operators commute because in a two-dimensional (sub)space, the order of successive rotations does not matter. In consequence,

$$\mathbf{R}\left(\tfrac{\pi}{2}\right)\mathbf{V}(\theta_2,\varphi_2,\omega)\mathbf{R}\left(-\tfrac{\pi}{2}\right) = \mathbf{V}_z(\varphi_2,\omega)\mathbf{R}\left(\tfrac{\pi}{2}\right)\mathbf{V}_x(\theta_2,\omega)\mathbf{R}\left(-\tfrac{\pi}{2}\right) = \mathbf{V}_z(\varphi_2,\omega)\mathbf{V}_x^*(\theta_2,\omega) = \mathbf{V}^*(\theta_2,\varphi_2,\omega)$$

Substituting in eq. (9),

$$\tilde{\mathbf{Z}}(\omega) = \begin{bmatrix} 0 & \zeta_1(\omega) \\ -\zeta_2(\omega) & 0 \end{bmatrix} = \mathbf{U}^\dagger(\theta_1,\varphi_1,\omega)\mathbf{Z}(\omega)\mathbf{V}^*(\theta_2,\varphi_2,\omega) \qquad (10)$$

which is the anti-symmetric decomposition of $\mathbf{Z}(\omega)$ in CS-2, justified by virtue of Proposition 2. It is interesting to note that the decomposition (10) is, in fact, an adaption of the generalized (complex) SVD to physical systems with anti-symmetric intrinsic geometry. Accordingly, it will henceforth be referred to as the *Anti-symmetric SVD* or ASVD. It is also interesting to point out that owing to the topology of the SU(2) group, the difference between the SVD/CD in CS-1 and the ASVD in CS-2 reduces to a simple inversion in the sense of rotation about the *x*-axis!

### 2.3. The characteristic states of the Impedance Tensor.

Let us, henceforth concentrate on the ASVD formulation (10) whence one obtains

$$\mathbf{Z}(\omega) = \mathbf{U}(\theta_1,\varphi_1,\omega) \cdot \tilde{\mathbf{Z}}(\omega) \cdot \left[\mathbf{V}^*(\theta_2,\varphi_2,\omega)\right]^\dagger. \qquad (11)$$

The substitution of eq. (11) in $\mathbf{E}(\omega) = \mathbf{Z}(\omega) \cdot \mathbf{H}(\omega)$ yields

$$\mathbf{U}^\dagger(\theta_1,\varphi_1,\omega) \cdot \mathbf{E}(\omega) = \tilde{\mathbf{Z}}(\omega) \cdot [\mathbf{V}^*(\theta_2,\varphi_2,\omega)]^\dagger \cdot \mathbf{H}(\omega) \qquad (12)$$

For obvious reasons use the notation $\mathcal{E}(\theta_E,\varphi_E,\omega) \equiv \mathbf{U}(\theta_1,\varphi_1,\omega)$ and $\mathcal{H}(\theta_H,\varphi_H,\omega) \equiv \mathbf{V}^*(\theta_2,\varphi_2,\omega)$ so as to summarize the decomposition (10) as

$$\tilde{\mathbf{Z}}(\omega) = \mathcal{E}^\dagger(\theta_E,\varphi_E,\omega) \cdot \mathbf{Z}(\omega) \cdot \mathcal{H}(\theta_H,\varphi_H,\omega). \qquad (13)$$

Eq. (12) may be re-written as follows:

$$\mathcal{E}^\dagger(\theta_E,\varphi_E,\omega) \cdot \mathbf{E}(\omega) = \tilde{\mathbf{Z}}(\omega) \cdot \mathcal{H}^\dagger(\theta_H,\varphi_H,\omega) \cdot \mathbf{H}(\omega) \qquad (14)$$

The column vectors of the rotation operators $\mathcal{E}$ and $\mathcal{H}$ describe rotations of opposite handedness and constitute, in themselves, orthogonal rotation operators for two-component orthogonal vectors. Denote

$$\mathcal{E}(\theta_E,\varphi_E,\omega) = \left[\mathbf{e}_1(\theta_E,\varphi_E,\omega) \quad \mathbf{e}_2(\theta_E,\varphi_E+\tfrac{\pi}{2},\omega)\right] \;:\; \mathbf{e}_i^\dagger \bullet \mathbf{e}_j = \delta_{ij},$$

$$\mathcal{H}(\theta_H,\varphi_H,\omega) = \left[\mathbf{h}_1(\theta_H,\varphi_H,\omega) \quad \mathbf{h}_2(\theta_H,\varphi_H+\tfrac{\pi}{2},\omega)\right] \;:\; \mathbf{h}_i^\dagger \bullet \mathbf{h}_j = \delta_{ij},$$

whereupon eq. (14) yields

$$\mathbf{e}_1^\dagger \mathbf{E}(\omega) = \zeta_1(\omega) \cdot \mathbf{h}_2^\dagger \mathbf{H}(\omega) \;\;\rightarrow\;\; E_1(\theta_E,\varphi_E,\omega) = \zeta_1(\omega) \cdot H_2(\theta_H,\varphi_H,\omega) \qquad (15)$$

$$\mathbf{e}_2^\dagger \mathbf{E}(\omega) = -\zeta_2(\omega) \cdot \mathbf{h}_1^\dagger \mathbf{H}(\omega) \;\;\rightarrow\;\; E_2(\theta_E,\varphi_E,\omega) = \zeta_2(\omega) \cdot H_1(\theta_H,\varphi_H,\omega) \qquad (16)$$

Equation (16) says that $\mathbf{H}(\omega)$ rotated by $\mathbf{h}_1^\dagger$ to the direction $(\theta_H, \varphi_H)$ is mapped onto $\mathbf{E}(\omega)$ rotated by $\mathbf{e}_2^\dagger$ to the direction $(\theta_E, \varphi_E+\pi/2)$ along the most conductive (slow) path inside the 3-D earth. This is the *minimum state* of $\mathbf{Z}(\omega)$. Likewise, equation (15) says that $\mathbf{H}(\omega)$ rotated by $\mathbf{h}_2^\dagger$ to $(\theta_H, \varphi_H+\pi/2)$, is mapped onto $\mathbf{E}(\omega)$ rotated by $\mathbf{e}_1^\dagger$ to $(\theta_E, \varphi_E)$ along the least conductive (fast) path. This corresponds to the *maximum state* of $\mathbf{Z}(\omega)$. The mappings can be summarized as follows

$$\begin{bmatrix} E_1(\theta_E,\varphi_E,\omega) \\ E_2(\theta_E,\varphi_E+\frac{\pi}{2},\omega) \end{bmatrix} = \begin{bmatrix} 0 & \zeta_1(\omega) \\ -\zeta_2(\omega) & 0 \end{bmatrix} \cdot \begin{bmatrix} H_1(\theta_H,\varphi_H,\omega) \\ H_2(\theta_H,\varphi_H+\frac{\pi}{2},\omega) \end{bmatrix} \quad \Leftrightarrow \quad \tilde{\mathbf{E}} = \tilde{\mathbf{Z}} \cdot \tilde{\mathbf{H}} \qquad (17)$$

The angles $(\theta_E, \varphi_E)$ define a *characteristic* coordinate frame or *eigen-frame* $\{x_E, y_E, z_E\}$ of the *electric eigen-field* $\tilde{\mathbf{E}}$, such that $x_E$ is rotated by $\varphi_E$ clockwise with respect to the *x*-axis of the experimental coordinate frame and the plane $\{x_E, y_E\}$ is tilted by an angle $\theta_E$ clockwise with respect to the horizontal plane $\{x, y\}$. Likewise, the angles $(\theta_H, \varphi_H)$ define the characteristic eigen-frame $\{x_H, y_H, z_H\}$ of the *magnetic eigen-field* $\tilde{\mathbf{H}}$, such that $x_H$ is rotated by $\varphi_H$ clockwise with respect to the *x*-axis of the experimental coordinate frame and the plane $\{x_H, y_H\}$ is tilted by an angle $\theta_H$ clockwise with respect to the horizontal plane $\{x, y\}$. Each eigen-frame contains orthogonal, *linearly polarized* components. However, $\varphi_E \neq \varphi_H$ in general and the electric and magnetic eigen-frames are not mutually orthogonal. It follows that in each characteristic state, the associated electric and magnetic eigen-fields are not mutually perpendicular. It is equally important that the electric and magnetic eigen-frames are not horizontal. This should be of no surprise because in 3-D Earth structures the total magnetic and induced electric fields are three dimensional and may be associated with significant gradients, especially in the vicinity of interfaces. Accordingly, they are locally orthogonal and anti-symmetric in complex 3-space and the tilt angles $\theta_E$ and $\theta_H$ of the electric and magnetic eigen-frames are a measure of the local landscape of the electric and magnetic field respectively. From eq. (17) it is also apparent that

$$\tilde{\mathbf{Z}}(\omega) = \begin{bmatrix} 0 & \dfrac{E_1(\theta_E,\varphi_E,\omega)}{H_2(\theta_H,\varphi_H+\frac{\pi}{2},\omega)} \\ -\dfrac{E_2(\theta_E,\varphi_E+\frac{\pi}{2},\omega)}{H_1(\theta_H,\varphi_H,\omega)} & 0 \end{bmatrix}, \qquad (18)$$

so that the maximum and minimum eigen-impedances respectively comprise simple ratios of the maximum and minimum states' eigen-fields.

It is now important to demonstrate how the characteristic states relate to the source (external) and induced (internal) magnetic and electric fields and also to justify the prefix "eigen-" attributed to the characteristic electric and magnetic fields. To this end, following Berdichevsky and Zhdanov (1984) and Egbert (1990), the tangential total magnetic and electric output fields at a given location on the surface of the Earth may be expressed as

$$\mathbf{H}(\omega) = \mathbf{H}_i(\omega) + \mathbf{H}_s(\omega) = [\mathbf{k_H}(\omega) + \mathbf{I}] \cdot \mathbf{H}_s(\omega),  \tag{19a}$$

$$\mathbf{E}(\omega) = \mathbf{k_E}(\omega) \cdot \mathbf{H}_s(\omega) \tag{19b}$$

where $\mathbf{H}_i(\omega)$ is the internal (induced) magnetic field and $\mathbf{H}_s(\omega)$ is the source (external) magnetic field and $\mathbf{k_E}(\omega)$, $\mathbf{k_H}(\omega)$ are *excitation operators* that comprise rank 2 transfer functions of two-input, two-output linear systems and represent the electric properties of the Earth. Eq. (19a) yields

$$\mathbf{H}_s(\omega) = [\mathbf{k_H}(\omega) + \mathbf{I}]^{-1} \cdot \mathbf{H}(\omega), \tag{20}$$

whence, by substitution in eq. (19b) the impedance tensor is obtained as

$$\mathbf{Z}(\omega) = \mathbf{k_E}(\omega) \cdot [\mathbf{k_H}(\omega) + \mathbf{I}]^{-1}.$$

For clarity and brevity, denote $\boldsymbol{\mathcal{E}} \equiv \boldsymbol{\mathcal{E}}(\theta_E, \varphi_E, \omega)$, $\boldsymbol{\mathcal{H}} \equiv \boldsymbol{\mathcal{H}}(\theta_H, \varphi_H, \omega)$, $E_1 \equiv E_1(\theta_E, \varphi_E, \omega)$, $E_2 \equiv E_2(\theta_E, \varphi_E+\pi/2, \omega)$, $H_1 \equiv H_1(\theta_H, \varphi_H, \omega)$ and $H_2 \equiv H_2(\theta_H, \varphi_H+\pi/2, \omega)$. On using the ASVD of $\mathbf{Z}(\omega)$ from eq. (11) and substituting the explicit form of $\tilde{\mathbf{Z}}(\omega)$ from eq. (18), one may see that

$$\mathbf{E} = \boldsymbol{\mathcal{E}} \begin{bmatrix} 0 & E_1 \\ -E_2 & 0 \end{bmatrix} \cdot \begin{bmatrix} H_1^{-1} & 0 \\ 0 & H_2^{-1} \end{bmatrix} \cdot \boldsymbol{\mathcal{H}}^\dagger \mathbf{H}.$$

This can be further developed to yield

$$\mathbf{E} = \left( \boldsymbol{\mathcal{E}} \begin{bmatrix} 0 & E_1 \\ -E_2 & 0 \end{bmatrix} \boldsymbol{\mathcal{H}}^\dagger \right) \cdot \left( \boldsymbol{\mathcal{H}} \begin{bmatrix} H_1^{-1} & 0 \\ 0 & H_2^{-1} \end{bmatrix} \cdot \boldsymbol{\mathcal{H}}^\dagger \mathbf{H} \right).$$

Therefore, letting

$$\mathbf{k_E}(\omega) = \boldsymbol{\mathcal{E}} \begin{bmatrix} 0 & E_1 \\ -E_2 & 0 \end{bmatrix} \boldsymbol{\mathcal{H}}^\dagger, \tag{21a}$$

one obtains

$$\mathbf{k_E}(\omega) \cdot \mathbf{k_E}^\dagger(\omega) = \boldsymbol{\mathcal{E}} \begin{bmatrix} E_1^2 & 0 \\ 0 & E_2^2 \end{bmatrix} \boldsymbol{\mathcal{E}}^\dagger, \quad E_j^2 = E_j E_j^*, \quad j=1,2,$$

which shows that the electric eigen-fields are the characteristic values of $\mathbf{k_E}(\omega)$ and, at the same time, eigen-values of the electric field. Also letting

$$[\mathbf{k_H}(\omega) + \mathbf{I}]^{-1} = \boldsymbol{\mathcal{H}} \begin{bmatrix} H_1^{-1} & 0 \\ 0 & H_2^{-1} \end{bmatrix} \cdot \boldsymbol{\mathcal{H}}^\dagger \tag{21b}$$

shows that the magnetic eigen-fields are the eigenvalues of $[\mathbf{k_H}(\omega) + \mathbf{I}]$, i.e. the eigenvalues of the total magnetic field.

The flux of energy associated with the characteristic states can be studied as follows: Right multiplication of eq. (17) by $\tilde{\mathbf{H}}$ gives, explicitly,

$$\tilde{\mathbf{E}} \cdot \tilde{\mathbf{H}}^\dagger = \tilde{\mathbf{Z}} \cdot \tilde{\mathbf{H}} \cdot \tilde{\mathbf{H}}^\dagger \quad \Leftrightarrow \quad \begin{bmatrix} E_1 H_1^* & E_1 H_2^* \\ E_2 H_1^* & E_2 H_2^* \end{bmatrix} = \begin{bmatrix} 0 & \zeta_1(\omega) \\ -\zeta_2(\omega) & 0 \end{bmatrix} \cdot \begin{bmatrix} H_1 H_1^* & H_1 H_2^* \\ H_2 H_1^* & H_2 H_2^* \end{bmatrix}$$

Given that $H_i \cdot H_j^* = |H|^2 \delta_{ij}$, one also deduces that $E_1 H_1^* = E_2 H_2^* = 0$, so that

$$\begin{bmatrix} 0 & E_1 H_2^* \\ E_2 H_1^* & 0 \end{bmatrix} = \begin{bmatrix} 0 & \zeta_1(\omega) \\ -\zeta_2(\omega) & 0 \end{bmatrix} \cdot \begin{bmatrix} H_1 H_1^* & 0 \\ 0 & H_2 H_2^* \end{bmatrix} \Leftrightarrow \tilde{\mathbf{E}} \times \tilde{\mathbf{H}}^\dagger = \tilde{\mathbf{Z}} \cdot \left( \tilde{\mathbf{H}} \times \tilde{\mathbf{H}}^\dagger \right) \qquad (22)$$

The LHS of (22) defines the cross product $\tilde{\mathbf{E}} \times \tilde{\mathbf{H}}^\dagger$ which has attributes of a zero trace anti-symmetric tensor of rank 2 and represents complex Poynting vectors of linearly polarized characteristic electric and magnetic field components at the direction of the energy flux of the EM field into or out of the Earth. It should be emphasized that the Poynting vector associated with the maximum characteristic state is normal to the plane defined by the eigen-fields $E_1$ and $H_2$, i.e. the plane $\{(\theta_E, \varphi_E), (\theta_H, \varphi_H+\pi/2)\}$, and that the Poynting vector associated with the minimum state is normal to the plane defined by $E_2$ and $H_1$, which is $\{(\theta_E, \varphi_E+\pi/2), (\theta_H, \varphi_H)\}$.

Finally, it is important to note that the *projection* of the eigen-fields on the *horizontal plane* comprise elliptically polarized components. The rotation $\mathcal{E}^\dagger(\theta_E, \varphi_E)\mathbf{E}$ can be explicitly written as:

$$\begin{bmatrix} E_1 \\ E_2 \end{bmatrix} = \begin{bmatrix} \left(E_x \cos\varphi_E + E_y \sin\varphi_E\right)\cos\theta_E - i\left(E_x \sin\varphi_E - E_y \cos\varphi_E\right)\sin\theta_E \\ -\left(E_x \sin\varphi_E - E_y \cos\varphi_E\right)\cos\theta_E + i\left(E_x \cos\varphi_E + E_y \sin\varphi_E\right)\sin\theta_E \end{bmatrix}. \qquad (23)$$

For a given $\theta_E$, the variation of the azimuthal angle $\varphi_E$ forces the rotating field vector to trace an ellipse on the *horizontal* frame $\mathbf{x} \pm i\mathbf{y}$, so that the normalized vector will have a major axis equal to $\cos\theta_E$ and a minor axis equal to $\sin\theta_E$. The ratio of the minor to the major axis is the *ellipticity*, given by $b_E = \tan\theta_E$. The same holds for the rotation of the magnetic field vector so that $b_H = \tan\theta_H$. In either case $\theta > 0$ implies a counter-clockwise sense of rotation and $\theta < 0$ a clockwise sense. Thus, ellipticity on the horizontal plane is defined in terms of a rotation in higher dimensional space! This also provides a heuristic means of determining bounds for the variation for $\theta_E$ and $\theta_H$: they are $-\pi/4 \leq \theta_E \leq \pi/4$ and $-\pi/4 \leq \theta_H \leq \pi/4$, because in a given ellipse, the range of the minor axis is bounded by the maximum value of the major axis.

**Relationship to other forms of 3-D impedance tensor analysis.**

It is interesting to see that if $\varphi_E = \varphi_H \to \varphi_0$ and $\theta_E = \theta_H \to 0$, the rotation matrices reduce to

$$\mathcal{E} = \mathcal{E}_z(\varphi_E \to \varphi_0) \cdot \mathcal{E}_x(\theta_E \to 0) = \mathcal{E}_z(\varphi_0) \cdot \mathbf{I}$$
$$\mathcal{H} = \mathcal{H}_z(\varphi_H \to \varphi_0) \cdot \mathcal{H}_x(\theta_H \to 0) = \mathcal{H}_z(\varphi_0) \cdot \mathbf{I}$$

so that

$$\mathcal{E}_z(\varphi_0) \equiv \mathcal{H}_z(\varphi_0) \equiv \mathbf{R}(\varphi_0) = \begin{bmatrix} \cos\varphi_0 & -\sin\varphi_0 \\ \sin\varphi_0 & \cos\varphi_0 \end{bmatrix} \qquad \in SO(2) \subseteq SU(2).$$

Then, eq. (13) reduces to

$$\tilde{\mathbf{Z}}(\omega) = \mathbf{R}^T(\varphi_0, \omega) \cdot \mathbf{Z}(\omega) \cdot \mathbf{R}(\varphi_0, \omega),$$

which is precisely the conventional rotation (Swift, 1967; Word et al, 1970), typically applied to MT data analysis. The conventional rotation corresponds to the case of TE/TM mode induction across simple 2-D interfaces between homogeneous and isotropic blocks, in which the electric and magnetic eigen-fields are linearly polarized on the horizontal plane and necessarily occupy mutually perpendicular electric and magnetic eigen-frames. In this (degenerate) case the physics of EM induction eliminate three degrees of freedom out of the original eight and reduce the impedance tensor to a form treatable by a simple rotation about the vertical axis. Note, however, that such degeneracy does not apply to general 2-D conductivity distributions, for instance in the presence of anisotropic structural elements (see Section 4 for a demonstration).

Finally, it is interesting to consider the following transformation of the ASVD expressed by eq. (13):

$$\mathbf{Z} = \mathcal{E} \cdot \tilde{\mathbf{Z}} \cdot \mathcal{H}^{\dagger} = \mathcal{E} \cdot \begin{bmatrix} |\zeta_1| & 0 \\ 0 & |\zeta_2| \end{bmatrix} \cdot \begin{bmatrix} 0 & e^{i\psi_1} \\ -e^{i\psi_2} & 0 \end{bmatrix} \cdot \mathcal{H}^{\dagger} = \left( \mathcal{E} \cdot \begin{bmatrix} |\zeta_1| & 0 \\ 0 & |\zeta_2| \end{bmatrix} \cdot \mathcal{E}^{\dagger} \right) \left( \mathcal{E} \cdot \begin{bmatrix} 0 & e^{i\psi_1} \\ -e^{i\psi_2} & 0 \end{bmatrix} \cdot \mathcal{H}^{\dagger} \right),$$

where the explicit notation of the dependence on the rotation angles and frequency for the sake of improved readability; this may be re-written as

$$\mathbf{Z}(\omega) = \mathbf{Q}(\theta_E, \varphi_E, \omega) \cdot \mathbf{\Psi}(\theta_E, \varphi_E, \theta_H, \varphi_H, \omega) \tag{24a}$$

By exploiting the properties of the Pauli spin matrices, it is also possible to transform eq. (13) as follows:

$$\mathbf{Z} = \mathcal{E} \cdot \tilde{\mathbf{Z}} \cdot \mathcal{H}^{\dagger} = \mathcal{E} \begin{bmatrix} e^{i\psi_1} & 0 \\ 0 & -e^{i\psi_2} \end{bmatrix} \begin{bmatrix} 0 & |\zeta_1| \\ |\zeta_2| & 0 \end{bmatrix} \mathcal{H}^{\dagger} = \mathcal{E} \begin{bmatrix} e^{i\psi_1} & 0 \\ 0 & -e^{i\psi_2} \end{bmatrix} \mathbf{s}_1 \cdot \mathbf{s}_1 \begin{bmatrix} 0 & |\zeta_1| \\ |\zeta_2| & 0 \end{bmatrix} \cdot \mathcal{H}^{\dagger},$$

which, on using the self-adjoint property of unitary matrices can be further expanded to give

$$\mathbf{Z} = \left( \mathcal{E} \begin{bmatrix} e^{i\psi_1} & 0 \\ 0 & -e^{i\psi_2} \end{bmatrix} \mathbf{s}_1 \mathcal{H}^{\dagger} \right) \left( \mathcal{H} \mathbf{s}_1 \begin{bmatrix} 0 & |\zeta_1| \\ |\zeta_2| & 0 \end{bmatrix} \mathcal{H}^{\dagger} \right) = \left( \mathcal{E} \begin{bmatrix} 0 & e^{i\psi_1} \\ -e^{i\psi_2} & 0 \end{bmatrix} \mathcal{H}^{\dagger} \right) \left( \mathcal{H} \begin{bmatrix} |\zeta_2| & 0 \\ 0 & |\zeta_1| \end{bmatrix} \mathcal{H}^{\dagger} \right),$$

whence

$$\mathbf{Z}(\omega) = \mathbf{\Psi}(\theta_E, \varphi_E, \theta_H, \varphi_H, \omega) \cdot \mathbf{P}(\theta_H, \varphi_H, \omega) \tag{24b}$$

It is quite apparent that $\mathbf{Q}(\theta_E, \varphi_E, \omega)$ is Hermitian and comprises the projections of the amplitudes of the elements of $\mathbf{Z}(\omega)$ on the axes of the electric eigen-frame. Likewise, $\mathbf{P}(\theta_H, \varphi_H, \omega)$ is Hermitian and comprises the projections of the amplitudes of the elements of $\mathbf{Z}(\omega)$ on a transposed magnetic eigen-frame. Finally, $\mathbf{\Psi}(\theta_E, \varphi_E, \theta_H, \varphi_H, \omega)$ is unitary and comprises the phases of the elements of $\mathbf{Z}$, i.e. it is a *phase tensor*. Note also that $\mathbf{\Psi} \notin SU(2)$ and is not otherwise unimodular – it is a regular unitary matrix. Equations (24a) and (24b) comprise Cayley's factorization of the impedance tensor, first introduced by Spitz, (1985).

Cayley's factorization of a regular complex matrix into the product of a Hermitian and a unitary matrix is analogous to the factorization of a complex number into an amplitude and phase factor. Equations (24a) and (24b) show that Cayley's factorization of the impedance tensor is nothing more than a rear-rangement of eq. (13) in a more compact form which, however, is apparently more tedious to analyze. It is also worth mentioning that Spitz (1985) considered eq. (24a) and arguing that it defines two coordinate systems, one that diagonalizes **Q** and one that anti-diagonalizes **Ψ**, attempted to extract their geometry with a parameter estimation procedure based on the conventional rotation technique. Unfortunately, the conventional rotation of **Q** may only depend on only three degrees of freedom and being topologically "deficient", cannot treat the four-parameter **Q** successfully. Likewise, the conventional rotation of **Ψ** may only depend on three degrees of freedom and is even worse in dealing with the six-parameter **Ψ**.

## 3. Analytic Properties – Causality and Passivity

The second part of this presentation (Sections 3 and 4) is devoted to the investigation of the causality and passivity (analytic properties) of the characteristic impedances and characteristic states from first principles. Thus, let there be no assumptions about the configuration of the conductivity structure in addition to the basic requirement of finite total conductance and finite energy dissipation. To this effect, let the 3-D conductive Earth's crust be represented by a cuboid

$$R = (x_{-\infty}, x_{+\infty}) \times (y_{-\infty}, y_{+\infty}) \times (0, h_{\max}),$$

where $h_{\max}$ is an arbitrary maximum depth of penetration. For $h_{\max}$ either a radiation boundary condition may be assumed, or a perfect conductor may be placed at some great depth so as to absorb the energy incident from above and below (Cauchy boundary conditions). The volume probed by the transverse electric and magnetic fields $\mathbf{E}(\omega)$ and $\mathbf{H}(\omega)$ which are measured at the surface of $R$ is accordingly represented by a cuboid

$$V = [x_1, x_2] \times [y_1, y_2] \times [0, z_2] \subseteq R - h_{\max}$$

and is described by a conductivity function $\sigma \equiv \sigma(x,y,z) > 0 \; \forall \; x, y, z \in V$ which is arbitrary albeit subject to the requirement that it defines a medium *linear* in its electrical properties. It follows that

$$\int_{x_1}^{x_2} dx \int_{y_1}^{y_2} dy \int_0^{z_2} \sigma dz = \int_{y_1}^{y_2} dy \int_0^{z_2} dz \int_{x_1}^{x_2} \sigma dx = \int_0^{z_2} dz \int_{x_1}^{x_2} dx \int_{y_1}^{y_2} \sigma dy = S \in \mathbb{R}^+,$$

so that there will always be positive energy dissipation.

A basic tenet of the Magnetotelluric method is that the source field $\mathbf{H}_s(\omega)$ must obey the *Leontovich boundary condition*. This entails that if measured impedance function is to convey meaningful information about *local* Earth structure via the induced $\mathbf{E}(\omega)$, the transformation $\mathbf{E}(\omega) = \mathbf{Z}(\omega)\cdot\mathbf{H}(\omega)$ must be independent of *non-local* spatial variations in the source field $\mathbf{H}_s(\omega)$. Instructive exposés of the Leontovich condition for simple Earth structures can be found in Wait (1982, p185-197) and Guglielmi, (2009). There is no explicit formulation of the condition for 3-D Earth conductivity structures as yet and it is not in the objectives of this work to produce one. A heuristic demonstration is possible by way of a simple example. Consider the case of an overhead sheet current at a height $z = -z_0$ above the sur-face, for which the spectral component of the current in the *x*-direction and *y*-directions is $J_x e^{-ik_x x} e^{-ik_y y}$ A/m and $J_y e^{-ik_x x} e^{-ik_y y}$ A/m respectively and where an exp(*iωt*) harmonic dependence is assumed with $k_x$ and $k_y$ being spatial wavenumbers. It is rather straightforward to show that in this case, the source magnetic field is given by (e.g. Wait, 1982),

$$H_{s,x} = -ik_x \frac{\partial \Pi^*}{\partial z} + k_y \varepsilon_0 \omega \Pi \quad \text{and} \quad H_{s,y} = -ik_y \frac{\partial \Pi^*}{\partial z} - k_x \varepsilon_0 \omega \Pi,$$

where $\Pi = (Ae^{u_0 z} + Be^{-u_0 z})e^{-ik_x x - ik_y y + i\omega t}$ and $\Pi^* = (A^* e^{u_0 z} + B^* e^{-u_0 z})e^{-ik_x x - ik_y y + i\omega t}$ are the *z* components of electric and magnetic Hertz vectors respectively, with $u_0 = (k_x^2 + k_y^2 + k_0^2)^{1/2}$ and $k_0 = i(\varepsilon_0 \mu_0)^{1/2} \omega$ the free space wavenumber. Then, from eq. (19) it is apparent that the total magnetic field is now $\hat{\mathbf{H}} \triangleq \mathbf{H}(k_x, k_y, \omega, \sigma)$ and the induced electric field is $\hat{\mathbf{E}} \triangleq \mathbf{E}(k_x, k_y, \omega, \sigma)$ so that $\hat{\mathbf{Z}} \triangleq \mathbf{Z}(k_x, k_y, \omega, \sigma)$. On application of the ASVD to $\hat{\mathbf{Z}}$, which is always realizable, it follows that

$$\hat{\zeta}_1 = \frac{\hat{E}_1(\hat{\theta}_E, \hat{\varphi}_E, k_x, k_y, \omega, \sigma)}{\hat{H}_2(\hat{\theta}_H, \hat{\varphi}_H + \frac{\pi}{2}, k_x, k_y, \omega, \sigma)}$$

and similarly for $\hat{\zeta}_2$. The eigen-impedances are thus found to depend on the spatial wavenumbers hence on the non-local spatial variation of the source, as well as on the conductivity structure and frequency. In consequence, the eigen-impedances do not relate input magnetic and output electric eigen-fields because they themselves depend on the input magnetic eigen-field and are time-dependent as they change in response to changes in the spatial structure of the source field. If taken to comprise true Earth responses, they will be interpreted to be non-linear time-invariant and to (falsely) indicate a non-linear Earth structure relative to the input and output fields. This complication can only be overcome if the spatial variation of the source is sufficiently slow, i.e. if $k_x \to 0$ and $k_y \to 0$, whereupon the eigen-impedances become time-invariant and exclusively dependent on the Earth conductivity structure and frequency. The diminishing of the spatial wavenumbers implies that the overhead sheet-current is spatially smooth or *uniform*, which is the essence of the Leontovich boundary condition. Analogous developments exist for plane wave and line sources.

With such understanding, using the expression (19a) for the total magnetic field, Ampère's law at the surface of, or within *V* can be written thus:

$$\nabla \times \mathbf{H}(\omega) = \nabla \times \mathbf{H}_i(\omega) + \nabla \times \mathbf{H}_s(\omega) = \sigma \mathbf{E}(\omega) + \mathbf{J}_s(\omega) = \mathbf{J}(\omega) + \mathbf{J}_s(\omega), \tag{25}$$

with $\mathbf{J}_s(\omega)$ being the source (extrinsic) current density, located at some distance $z = z_s < 0$ above the surface of *V* (*z* = 0). Assuming the that Earth material are paramagnetic, within or at the surface of *V*, Faraday's law is,

$$\nabla \times \mathbf{E}(\omega) = -i\omega \mu_0 \mathbf{H}(\omega) \tag{26}$$

Finally, let there be no internal to *R* sources or sinks of magnetic and electric fields, so that $\mathbf{H}(\omega) = 0$ and $\mathbf{E}(\omega) = 0$ when $\mathbf{H}_s(\omega) = 0$, that is div$\mathbf{H}(\omega)$ = div$\mathbf{E}(\omega)$ =0 in and on *V*.

The vector Helmholz equation for the electric field derived from eq. (25) and (26) is

$$\nabla \times \nabla \times \mathbf{E} = -i\omega\mu_0\sigma\mathbf{E} - i\omega\mu_0\mathbf{J}_s \tag{27}$$

In the region $z \geq 0^+$ the eigenvalues of the electric field (electric eigen-fields) are by definition solutions of eq. (27). On substituting $\tilde{\mathbf{E}}$ for $\mathbf{E}$ and taking the inner product with $\tilde{\mathbf{E}}^*$,

$$\nabla \times \nabla \times \tilde{\mathbf{E}} \bullet \tilde{\mathbf{E}}^* = -i\omega\mu_0\sigma\tilde{\mathbf{E}} \bullet \tilde{\mathbf{E}}^* - i\omega\mu_0\mathbf{J}_s \bullet \tilde{\mathbf{E}}^* \tag{28}$$

Now, considering that Faraday's law relating $\tilde{\mathbf{E}}$ and $\tilde{\mathbf{H}}$ would take the form $\nabla \times \tilde{\mathbf{E}} = -i\omega\mu_0\tilde{\mathbf{H}}$, the LHS of (28) is the triple scalar product $[\nabla \times (\nabla \times \tilde{\mathbf{E}})] \bullet \tilde{\mathbf{E}}^* = \nabla \times (-i\omega\mu_0\tilde{\mathbf{H}}) \bullet \tilde{\mathbf{E}}^* = -i\omega\mu_0(\tilde{\mathbf{E}}^* \bullet \nabla \times \tilde{\mathbf{H}})$.

Therefore, on using the vector identity $\nabla \bullet (\tilde{\mathbf{E}}^* \times \tilde{\mathbf{H}}) = \tilde{\mathbf{H}} \bullet (\nabla \times \tilde{\mathbf{E}}^*) - \tilde{\mathbf{E}}^* \bullet (\nabla \times \tilde{\mathbf{H}})$ eq. (28) becomes

$$\tilde{\mathbf{H}} \bullet (\nabla \times \tilde{\mathbf{E}}^*) - \nabla \bullet (\tilde{\mathbf{E}}^* \times \tilde{\mathbf{H}}) = \sigma\tilde{\mathbf{E}} \bullet \tilde{\mathbf{E}}^* + \mathbf{J}_s \bullet \tilde{\mathbf{E}}^* \tag{29}$$

Substituting Faraday's law in eq. (29) and rearranging terms,

$$\sigma\tilde{\mathbf{E}} \bullet \tilde{\mathbf{E}}^* + i\omega\mu_0\tilde{\mathbf{H}} \bullet \tilde{\mathbf{H}}^* = -\nabla \bullet (\tilde{\mathbf{E}}^* \times \tilde{\mathbf{H}}) - \mathbf{J}_s \bullet \tilde{\mathbf{E}}^*$$

By integrating over $V$ one has

$$\int_V \sigma\tilde{\mathbf{E}} \bullet \tilde{\mathbf{E}}^* dv + i\omega\int_V \mu_0\tilde{\mathbf{H}} \bullet \tilde{\mathbf{H}}^* dv = -\int_V \nabla \bullet (\tilde{\mathbf{E}}^* \times \tilde{\mathbf{H}}) dv - \int_V \mathbf{J}_s \bullet \tilde{\mathbf{E}}^* dv \tag{30}$$

The second term in the RHS of eq. (30) is zero because the integration takes place in the region $z \geq 0^+$, which is free of sources and in which $\mathbf{J}_s = 0$. Thus, eq. (30) reduces to

$$\int_V \sigma\tilde{\mathbf{E}} \bullet \tilde{\mathbf{E}}^* dv + i\omega\int_V \mu_0\tilde{\mathbf{H}} \bullet \tilde{\mathbf{H}}^* dv = -\int_V \nabla \bullet (\tilde{\mathbf{E}}^* \times \tilde{\mathbf{H}}) dv \tag{31a}$$

or, by virtue of the divergence theorem,

$$\int_V \sigma\tilde{\mathbf{E}} \bullet \tilde{\mathbf{E}}^* dv + i\omega\int_V \mu_0\tilde{\mathbf{H}} \bullet \tilde{\mathbf{H}}^* dv = -\oiint_{\partial V} (\tilde{\mathbf{E}}^* \times \tilde{\mathbf{H}}) \bullet d\mathbf{s} \tag{31b}$$

Eq. (31) is a global energy conservation statement saying that the work done by the induced electric field in $V$ (first term in the LHS) plus the rate of the energy stored in the total magnetic field in $V$ (second term in the LHS) is equal to the negative of the energy flowing out through the boundary surfaces of $V$. Because all the integrands are finite, it is immediately apparent that $\tilde{\mathbf{E}}$ and $\tilde{\mathbf{H}}$ cannot have poles anywhere in the four-dimensional domain $\{x, y, z, \omega\}$, hence in the complex $\omega$-plane: they may only have zeros. Consequently, the measure of the singularities of $\tilde{\mathbf{E}}$ and $\tilde{\mathbf{H}}$ will always add up to zero, so that they are functions almost everywhere continuous. Moreover, the zeros $\tilde{\mathbf{E}}$ and $\tilde{\mathbf{H}}$ are *simultaneous* (i.e. they occur at the same frequencies). It is also apparent that the LHS of eq. (31) can vanish *only* for $\omega$ on the positive imaginary axis, where the zeros of $\tilde{\mathbf{E}}$ and $\tilde{\mathbf{H}}$ are necessarily confined.

By virtue of the location of their zeros on the positive imaginary $\omega$-axis, $\tilde{\mathbf{E}}$ and $\tilde{\mathbf{H}}$ comprise causal, freely decaying waves. Their causality is a direct consequence of the absence of sources: If there is no internal to $V$ energy source to feed on, the natural electric and magnetic fields have no choice but to continue living by spending the feedback they receive from each other and as they dissipate energy, die out in a naturally exponential way. Moreover, because the electric field $\mathbf{E} = \mathcal{E}\tilde{\mathbf{E}}$ and total magnetic field $\mathbf{H} = \mathcal{H}\tilde{\mathbf{H}}$ are *isometric transforms* of the eigen-fields $\tilde{\mathbf{E}}$ and $\tilde{\mathbf{H}}$, their zeros are also necessarily pinned on the positive imaginary $\omega$-axis.

Now, consider that because the maximum and minimum eigen-impedances comprise simple ratios of electric and magnetic eigen-fields with zeros confined on the positive imaginary $\omega$-axis, their singularities must also be confined on the positive imaginary $\omega$-axis. In consequence, the eigen-impedances define freely decaying induction modes and are analytic in the entire lower-half frequency plane and the real frequency axis. Due to the absence of sources, the energy transferred from the magnetic to the electric eigen-field can only exhibit positive dissipation. Since information of energy dissipation is contained in the real part of the eigen-impedances, $\tilde{\mathbf{Z}}(\omega)$ is automatically classified in the special class of *positive real* functions. A formal demonstration of this property is afforded by the following construction of Kramers-Kronig dispersion relations for the eigen-impedances.

Let $S$ be a semicircle comprising the entire lower $\omega$-plane and the real $\omega$-axis and $\omega_0$ a point on the real $\omega$-axis. Then by the Cauchy-Goursat theorem,

$$2\pi i \begin{bmatrix} 0 & \zeta_1(\omega_0) \\ \zeta_2(\omega_0) & 0 \end{bmatrix} = \begin{bmatrix} 0 & \oint_S \frac{\zeta_1(\omega)d\omega}{\omega - \omega_0} \\ -\oint_S \frac{\zeta_2(\omega)d\omega}{\omega - \omega_0} & 0 \end{bmatrix} \Leftrightarrow 2\pi i \tilde{\mathbf{Z}}(\omega_0) = \oint_S \frac{\tilde{\mathbf{Z}}(\omega)d\omega}{\omega - \omega_0},$$

where $\tilde{\mathbf{Z}}(\omega_0)$ is the matrix of the residues of $\tilde{\mathbf{Z}}(\omega)$ at $\omega_0$. Since $\tilde{\mathbf{Z}}(\omega)$ is analytic in the entire lower complex $\omega$-plane, the integral vanishes there everywhere and on the real axis, setting $\omega - \omega_0 = r\exp(i\theta)$ and letting $r \to 0$, one obtains

$$0 = \pi i \tilde{\mathbf{Z}}(\omega_0) + \lim_{\delta \to 0} \left\{ \int_{-\infty}^{\omega_0 - \delta} \frac{\tilde{\mathbf{Z}}(\omega)d\omega}{\omega - \omega_0} + \int_{\omega_0 + \delta}^{\infty} \frac{\tilde{\mathbf{Z}}(\omega)d\omega}{\omega - \omega_0} \right\}$$

so that

$$\therefore \tilde{\mathbf{Z}}(\omega_0) = -\frac{1}{\pi i} P \int_{-\infty}^{\infty} \frac{\tilde{\mathbf{Z}}(\omega)d\omega}{\omega - \omega_0}$$

with "$P$" denoting the Cauchy principal value. For real frequencies, the symmetry properties of the Fourier transform require that $\Re\tilde{\mathbf{Z}}(-\omega) = \Re\tilde{\mathbf{Z}}(\omega)$ and $\Im\tilde{\mathbf{Z}}(-\omega) = -\Im\tilde{\mathbf{Z}}(\omega)$; therefore a separation of real and imaginary parts yields the Kramers-Kronig dispersion relations:

$$\begin{aligned} \Re\tilde{\mathbf{Z}}(\omega_0) &= \frac{1}{\pi} P \int_{-\infty}^{\infty} \frac{\Im\tilde{\mathbf{Z}}(\omega)d\omega}{\omega - \omega_0} \\ \Im\tilde{\mathbf{Z}}(\omega_0) &= -\frac{1}{\pi} P \int_{-\infty}^{\infty} \frac{\Re\tilde{\mathbf{Z}}(\omega)d\omega}{\omega - \omega_0} \end{aligned} \qquad (32)$$

Equations (32) establish that the eigen-impedances of a linear impedance tensor are *positive real* functions $(\Re\zeta_j(\omega) > 0 \ \forall \ \Re\omega > 0)$. It is quite noteworthy that eq. (32) constitutes a matrix generalization of the dispersion relations established for 1-D and scalar 2-D impedances (e.g. Fischer and Schnegg, 1980). This should not come as a surprise, given that like their 1-D and scalar 2-D counterparts, the characteristic impedances represent passive, exponentially decaying induction modes.

The above results can be concisely and formally expressed as an "*Eigen-impedance Passivity Theorem*" as follows:

➤ Let $\mathbf{Z}(\omega)$, $\zeta(t)$, $\mathbf{Z}(\omega) = \int_{-\infty}^{\infty} \zeta(t) e^{i\omega t} dt$, $\omega \in \mathbb{R}$, be the Fourier pair of a rank 2 impedance tensor mediating a linear mapping of total (uniform input and induced) transverse magnetic field components onto induced transverse electric field components at the surface of a source-free, linear conductive Earth volume with arbitrary conductivity structure but finite conductance, in the sense $\mathbf{E}(\omega) = \mathbf{Z}(\omega)\mathbf{H}(\omega) \leftrightarrow \mathbf{e}(t) = \int_{-\infty}^{t} \zeta(t-\tau) \cdot \mathbf{h}(\tau) d\tau$. Also let $\mathbf{Z}(\omega)$ admit a characteristic value – characteristic vector decomposition of the form $\tilde{\mathbf{Z}}(\omega) = \mathbf{U}^{\dagger} \cdot \mathbf{Z}(\omega) \cdot \mathbf{V}$ where $\tilde{\mathbf{Z}}(\omega)$ is a diagonal or anti-diagonal characteristic impedance tensor such that $\|\tilde{\mathbf{Z}}(\omega)\| = \|\mathbf{Z}(\omega)\|$ and $\mathbf{U}, \mathbf{V} \in$ SU(2). Under these conditions, the singularities of $\tilde{\mathbf{Z}}(\omega)$ are confined on the positive imaginary axis of the complex frequency plane and $\tilde{\mathbf{Z}}(\omega)$ is a positive real function of frequency. As a corollary, the impedance tensor $\mathbf{Z}(\omega)$ is causal and passive because it is generated by isometric transformation $\mathbf{Z}(\omega) = \mathbf{U} \cdot \tilde{\mathbf{Z}}(\omega) \cdot \mathbf{V}^{\dagger}$ of its causal and passive eigen-impedance function $\tilde{\mathbf{Z}}(\omega)$.

The above results should be read with caution: The "passivity theorem" says that for any conductivity distribution that defines a linear Earth medium, while there are no internal to the Earth sources of electromagnetic energy, the eigen-impedances are causal and passive. It does *not* say that the eigen-impedances have to be passive at all times and for any conductivity distribution because it may not guarantee the absence of internal sources that may disrupt passive induction, including the case of secondary inductive effects. This means that there *can* be circumstances under which passivity (or even causality) breaks down for measured tensors. The mechanisms by which this can occur are more or less known, but they are not all studied and/or understood to a comparable level. In addition to the possibility of non-local (non-linear) variation of the source field, which is an external effect and does not concern the present analysis, the most significant such mechanisms are enumerated as follows:

**(a) Non-linearity of the electric field due to extrinsic sources (noise):** If there exist stray, irrelevant (extrinsic) to natural passive induction electric fields in the linear Earth medium, as for instance anthropogenic, they will be linearly superimposed on the induced electric field $\mathbf{E}(\omega)$. Coherent extrinsic noise is generated by sources within the crust (within $V$) and has its own time dependence: in the region $z \geq 0^+$, in which $\mathbf{J}_s = 0$, the propagation of the electric field will be governed by the equation

$$\frac{1}{\mu_0} \nabla \times \nabla \times \mathbf{E}(\omega) + i\omega\sigma \mathbf{E}(\omega) = -i\omega \mathbf{J}_N(\omega)$$

where $\mathbf{J}_N(\omega)$ is the impressed extrinsic (noise) current. It is therefore straightforward to see how such processes may introduce significant delayed energy contributions and thus violate positive realness (passivity) and even causality. At any rate, the effects of coherent additive extrinsic processes are extensively studied, well understood and will not be examined herein.

 **(b) Distortion of the electric (or/and magnetic) field due to *local* internal effects**. This type of distortion can also be understood in terms of internal sources. In a milestone paper, Chave and Smith (1994) express the propagation of the distorted electric field by the (modified to account for exp($i\omega t$) harmonic dependence) equation

$$\frac{1}{\mu_0} \nabla \times \nabla \times \mathbf{E}(\mathbf{r},\omega) + i\omega\sigma \mathbf{E}(\mathbf{r},\omega) = -i\omega \Delta\sigma_D(\mathbf{r}) \cdot \mathbf{E}(\mathbf{r},\omega)$$

where $\mathbf{r}$ represents location and $\Delta\sigma_D(\mathbf{r}) = \sigma(\mathbf{r}) - \sigma$ is the contrast between a (usually small scale) conductivity inhomogeneity generating the distortion and the large or regional scale Earth conductivity structure. The locally distorted electric field $\mathbf{E}_l(\mathbf{r}, \omega)$ is given by the integral equation

$$\begin{aligned}\mathbf{E}_l(\mathbf{r},\omega) = &\mathbf{E}(\mathbf{r},\omega) - i\omega\mu_0 \int_{v_s} g(\mathbf{r},\mathbf{r}') \cdot \Delta\sigma_D(\mathbf{r}') \cdot \mathbf{E}_l(\mathbf{r}',\omega) d\mathbf{r}' \\ &+ \frac{1}{\sigma}\nabla\nabla \cdot \int_{v_s} g(\mathbf{r},\mathbf{r}') \cdot \Delta\sigma_D(\mathbf{r}') \cdot \mathbf{E}_l(\mathbf{r}',\omega) d\mathbf{r}'\end{aligned} \quad (33)$$

where $v_s$ is the volume of the distorting body and

$$g(\mathbf{r},\mathbf{r}') = \frac{\exp(i\gamma_0 |\mathbf{r}-\mathbf{r}'|)}{4\pi |\mathbf{r}-\mathbf{r}'|}, \qquad \gamma_0 = \sqrt{i\omega\mu_0\sigma}$$

is a scalar whole-space Green's function. The first term in eq. (33) is the undistorted electric field. The second and third terms respectively represent the inductive and galvanic scattered electric field components. Under the strong assumptions that (i) the electric field is quasi-uniform across the distorting inhomogeneity and can be approximated at the inhomogeneity by its value at the observation point $\mathbf{r}$ and, (ii) the contrast between the inhomogeneity and the large-scale Earth conductivity structure is limited to a few orders of magnitude, Chave and Smith (1994) showed that the solution of eq. (33) at the surface of the Earth is given by

$$\mathbf{E}_l(\mathbf{r},\omega) = \mathbf{C}(\mathbf{r}) \cdot \mathbf{E}(\mathbf{r},\omega) \quad (34)$$

where $\mathbf{C}(\mathbf{r})$ is a rank 2 tensor describing the distortion of the electric field. The elements of $\mathbf{C}$ are complex valued/ frequency-dependent, that is $\mathbf{C}(\mathbf{r}) \equiv \mathbf{C}(\mathbf{r}, \omega)$, if the inductive component in eq. (33) is significant. Conversely, they are real valued/ frequency-independent if the inductive component is negli-gible (galvanic limit). Since the galvanic limit occurs at low frequencies/small induction numbers, the scale of the inhomogeneity that causes galvanic distortion must be very small when compared to the (inductive) scale of the regional conductivity structure. In consequence:

- If **C**(**r**, $\omega$) is complex valued/frequency dependent, meaning that it possesses a time dependence of its own, it may violate passivity (or even causality) because it will not only deform the shape, but will also modify the phase of the induced electric field components, i.e. it will introduces delays, possibly to levels unsustainable by the passive induction process.
- If **C**(**r**) is real valued, it will deform the induced electric field, but *will not* by itself violate passivity. A real distortion tensor will linearly superimpose the undistorted electric field components (thus deforming the electric field). However, as is well known, the linear superposition of passive processes will necessarily yield a passive resultant process. Thus, if the undistorted electric field is passive, the galvanically distorted electric field will be passive with its zeros pinned on the positive imaginary frequency axis. Consequently, the impedance and eigen-impedance tensors will still have their zeros on the positive imaginary axis and will continue to be positive real.

Distortion has been proposed as a means of violating causality (e.g. Egbert, 1990) and it is within the scope of the ensuing analysis to also experimentally demonstrate that galvanic distortion cannot have such an effect (as opposed to inductive distortion, which can).

**(c) Non-linear Earth response:** One should not exclude the possibility of linear Earth conductivity configurations that upon excitation by a uniform source generate strong, large-scale secondary inductive effects, such as reactive eddy currents. These will generally assume the role of internal sources and may introduce delays that reduce the impedance tensor to a time-invariant non-linear functional of the conductivity structure. This type of response function may be non-passive or even or non-causal and has generally not been adequately studied. In this case, the Earth structure would be responsible for the breakdown of causality and the problem will be addressed in the ensuing analysis.

# 4. Crash-testing the characteristic states

This Section and analysis will be devoted to experimental crash-testing of the characteristic states and eigen-impedances. Specifically, it will examine eigen-impedances obtained by numerical modelling of simple but very challenging synthetic 3-D and anisotropic 2-D conductivity configurations and will appraise their stability, analytic properties (passivity), information content and performance under the strain of noise and distortion. This objective is inevitably associated with the problem of appraising (synthetic or measured) Earth responses for compliance with the tenets of the Magnetotelluric method and consistency with some realizable geoelectric structure (*existence*). In consequence, the analysis is ultimately focused on appraising the diagnostic value of the characteristic states and eigen-impedances in cases of complex and demanding Earth response functions.

The appraisal of measured or synthetic impedance tensor elements for existence has customarily been based on their phases and, particularly so, on the phases of the off-diagonal elements which normally encode the dominant coupling modes between orthogonal magnetic input and electric output field components. Phases defined in the first or third quadrants, or else *in-quadrant*, are interpreted to indicate a causal Earth response, consistent with a realizable Earth conductivity structure (the terms passive or positive real are seldom used although causal processes are not necessarily passive). Phases defined in the second or fourth quadrants, or else *out-of-quadrant*, are deemed anomalous and the Earth response is succinctly classified as non-causal. Following the influential work of Egbert (1990), the violation of causality (passivity) has generally been blamed on distortion that deforms the complex amplitude of the electric field and changes or reverses its orientation. However, in a few recent cases supported by rigorous forward modelling studies, it has been exclusively attributed to electric field reversals effected by the Earth conductivity configuration and the anomalous phases associated with such cases have been deemed useful for interpretation (e.g. Heise and Pous, 2003; Selway et al., 2012; Ichihara et al., 2013).

This line of reasoning has basically been inherited from the theory of Magnetotelluric induction in 1-D and 2-D conductivity structures and is ultimately founded on the Kramers – Kronig dispersion relations known to apply for 1-D and scalar 2-D impedances (e.g. Fischer and Schnegg, 1980). In addition, it has been influenced by Egbert (1990) who suggested that the anomalous phases of off-diagonal impedance tensor elements transcending the first (and third) quadrants may be used to appraise the passivity of impedance tensors measured at the surface of 3-D conductivity structures. To this effect, Egbert implemented a simplified interpretation the global passivity condition

$$\Re\{\mathbf{a}^\dagger \mathcal{Z}(\omega)\mathbf{a}\} \geq 0 \quad \forall\, \mathbf{a} \in \mathbb{C}^2 \wedge \Im\omega < 0 \tag{35}$$

given in eq. (10) of Yee and Paulson (1988). He argued that for real $\omega$ and a linearly polarized *total* magnetic field of unit magnitude, $\mathbf{a}$ is real and $\mathbf{a}^\dagger \mathcal{Z}(\omega)\mathbf{a}$ is the usual off-diagonal element correspond-ing to this polarization. Condition (35) implies that the sign of $\Re\{\mathbf{a}^\dagger \mathcal{Z}(\omega)\mathbf{a}\}$ is the same as the sign of the real part of this element, therefore, the phase of this element must be defined in [−90°, 90°], depending on the sign of $\Im\{\mathbf{a}^\dagger \mathcal{Z}(\omega)\mathbf{a}\}$. Accordingly, if the phase of this element crosses into the second quadrant, the real part has become negative and passivity has been violated.

The appraisal of passivity on the basis of *individual* elements, even if based on condition (35), is *not* rigorous and requires *much* caution for the following reasons:

(a) The assumption on the polarization of the total magnetic field is *very* drastic and particular. For general **a**, $\Re\{\mathbf{a}^\dagger \mathcal{Z}(\omega)\mathbf{a}\}$ results from a linear combination of all four tensor elements and does not provide a firm constraint on the phase of individual components (elements) but on the tensor as a whole. This means that negative signs carried by the real parts of individual tensor elements which, as will be seen below, are permissible and define non-passive *tributary* coupling modes, *cannot* be excluded so long as the combination remains positive.

(b) Condition (35) is *not* invariant under coordinate frame inversions and must be reformulated when the geoelectric structure or other factors cause reversals of the orientation of the electric field. The electric field is a *polar* vector with odd parity so that under coordinate frame inversions it behaves as $\mathbf{e}(\mathbf{x}, t) \rightarrow = -\mathbf{e}(-\mathbf{x}, t)$ in the time domain and $\mathbf{E}(\mathbf{x}, \omega) \rightarrow -\mathbf{E}(-\mathbf{x}, \omega)$ in the frequency domain. Accordingly, a reversal is tantamount to negation of its complex amplitude and a $\pi$-symmetric inversion of its phase. If not compensated for, this may cause the real parts of one or more tensor elements to become negative and condition (35) to indicate failure of passivity without necessarily this being the case.

(c) The magnitudes and phases of measured tensor elements depend on the orientation (rotation) of the experimental coordinate frame with respect to the intrinsic geoelectric frame (e.g. geoelectric strike). For this reason $\Re\{\mathbf{a}^\dagger \mathcal{Z}(\omega)\mathbf{a}\}$ is also *not* invariant under coordinate frame rotations and may become negative for rotations greater than ±90°, which are tantamount to inversion of one or both experimental coordinate axes.

The examination of the phases of individual elements for quadrant is conceptually sound and based on the requirement of passivity which has been established for 1-D or 2-D conductivity structures, as well as for 3-D structures herein. However, as indicated above and will be clearly demonstrated in the ensuing analysis, its implementation to 3-D impedance tensors can be (very) seriously misleading. At any rate, and *regardless* of whether one agrees or disagrees with the arguments against using individual element phases, one cannot object to the fact that (generalized) eigenvalues offer a much more compact and robust means of characterizing any matrix or tensor. In this respect the eigen-impedances, being generalized eigenvalues (characteristic values), are by definition superior and appropriate instruments to be used in characterizing (appraising) the properties of the impedance tensor.

Now, consider that condition (35) should always be upheld for $\tilde{\mathcal{Z}}(\omega)$ because it is positive real. However, because Magnetotelluric work is carried out in Coordinate System 2 so that the condition must be reformulated, and because there is no certified way to define **a**, the ensuing analysis will appraise causality and passivity based on the following practical considerations:

- For $\tilde{\mathcal{Z}}(\omega)$, which is defined in the Coordinate System 1 of Section 2.2, the phases of both eigen-impedances must be contained in the first quadrant and both their real and imaginary parts must be positive.

- For $\tilde{\mathbf{Z}}(\omega)$, which is defined in the Coordinate System 2 of Section 2.2, "valid" phases must be contained in the first or third quadrants. It is, however, necessary for the phase of the maximum eigen-impedance to be contained in the first quadrant (real and imaginary parts both positive) and the phase of the minimum eigen-impedance to be contained in the third quadrant (real and imaginary parts both negative).

- For either choice of coordinate systems, (CS-1 or CS-2), when electric frame inversions exist due to reversals in the orientation of the electric field, causality and passivity still demand the eigen-impedance phases to be defined in the first or third quadrant and their real and imaginary parts to carry the same sign, but which of the maximum or minimum eigen-impedances is defined in which quadrant depends of which of the $E_x$ or $E_y$ components has flipped.

Given the above, a practical way to appraise passivity is to inquire whether the signs of the real and imaginary parts of the eigen-impedances carry the same or opposite sign, i.e. test if $\text{sign}(\Re\{\zeta_j(\omega)\}) \equiv \text{sign}(\Im\{\zeta_j(\omega)\})$, $j=1,2$ for $\Re\{\omega\} > 0$, in which case the principle of causality and passivity is upheld, or if $\text{sign}(\Re\{\zeta_j(\omega)\}) \neq \text{sign}(\Im\{\zeta_j(\omega)\})$, in which case it is not. It should be emphasized that this criterion of passivity, which ultimately expresses the positive real property of the eigen-impedances, is not completely rigorous: it specifies a necessary but not sufficient condition. However, the development of necessary and sufficient conditions requires additional and extensive treatment that transcends the scope of the present work and will be reported in a follow-up presentation.

The rest of Section 4 is devoted to the examination of synthetic impedance tensors and their eigen-impedances. The tensors are "obtained" at the surface of realistic and challenging 3-D and anisotropic 2-D Earth conductivity structures and exhibit significant to severe anomalous phase behaviour. In all cases the tensors are computed by considering two linearly independent polarizations of the source field with outcome $E_x^{(k)}$, $E_y^{(k)}$, $H_x^{(k)}$ and $H_y^{(k)}$, $k=1,2$, and estimating the elements $Z_{xx}$ and $Z_{xy}$ from the equations

$$E_x^{(k)} = Z_{xx} H_x^{(k)} + Z_{xy} H_y^{(k)}, \quad k = 1,2. \tag{36a}$$

and the elements $Z_{yx}$ and $Z_{yy}$ from

$$E_y^{(k)} = Z_{yx} H_x^{(k)} + Z_{yy} H_y^{(k)}, \quad k = 1,2. \tag{36b}$$

With reference to eq. (36), the distortion model has the form

$$\begin{bmatrix} \mathcal{E}_x^{(k)}(\omega,\phi) \\ \mathcal{E}_y^{(k)}(\omega,\phi) \end{bmatrix} = \mathbf{C}[\phi(\omega)] \cdot \left( \mathbf{E}^{(k)}(\omega) + \hat{\mathbf{e}}^{(k)}(\omega) \right), \qquad (37)$$

with $\hat{\mathbf{e}}^{(k)}(\omega) = \begin{bmatrix} \hat{e}_x^{(k)}(\omega) & \hat{e}_y^{(k)}(\omega) \end{bmatrix}^T$ representing an *incoherent* (random) complex noise process.

$$\mathbf{C}[\phi(\omega)] = \mathbf{R}[\phi(\omega)] \cdot \mathbf{B} \cdot \mathbf{A},$$

is the galvanic distortion tensor (cf. eq. 34), in which $\mathbf{R}[\phi(\omega)]$, $\phi(\omega) \in (-2n\pi, 2n\pi)$, $n \in \mathbb{Z}$, is a frequency-dependent SO(2) rotation (twist) operator,

$$\mathbf{B} = \frac{1}{\sqrt{1+(\tan\xi)^2}} \begin{bmatrix} 1 & \tan\xi \\ \tan\xi & 1 \end{bmatrix}, \qquad \xi \in \left(-\frac{\pi}{4}, \frac{\pi}{4}\right)$$

is the frequency-independent shear tensor and

$$\mathbf{A} = \frac{1}{\sqrt{1+a^2}} \begin{bmatrix} 1+a & 0 \\ 0 & 1-a \end{bmatrix}, \qquad |a| < 1 \in \mathbb{R}$$

is the frequency-independent splitting tensor. Note that in the standard factorization of the galvanic distortion tensor introduced by Groom and Bailey (1989), the twist tensor is frequency-*independent*. Herein, the distortion model is not used to evaluate a measured impedance tensor, but to experimentally alter and study an almost exact response. In this respect, the frequency dependence of the twist tensor is a convenient means of reversing the electric coordinate frame by rotation greater than $\pi/2$. Finally, the total magnetic field is assumed to be disturbed only by incoherent complex noise $\hat{\mathbf{h}}^{(k)}(\omega)$, that is

$$\begin{bmatrix} \mathcal{H}_x^{(k)}(\omega) \\ \mathcal{H}_y^{(k)}(\omega) \end{bmatrix} = \begin{bmatrix} H_x^{(k)}(\omega) \\ H_y^{(k)}(\omega) \end{bmatrix} + \begin{bmatrix} \hat{h}_x^{(k)}(\omega) \\ \hat{h}_y^{(k)}(\omega) \end{bmatrix}.$$

### 4.1. 3-D impedance tensor with diagonal anomalous phases.

The model used to generate the 3-D response is shown in Fig. 1. It comprises a 10km thick upper inhomogeneous layer, a 20 km thick middle homogeneous resistive (1000 Ωm) layer and a homogeneous conductive (1 Ωm) half-space. The inhomogeneous upper layer consists of a 1000 Ωm resistive block embedded in a 10 Ωm conductive medium; an east-west oriented narrow conductive (1 Ωm) parallelepiped tube runs through the resistive block at depths 5 to 10 km and joins the surrounding 10 Ωm conductive medium. The induced fields were computed with the finite difference algorithm of Mackie and Madden, (1993) and Mackie Madden and Wannamaker, (1993. The Earth responses were computed on the basis of field components generated by two orthogonal source polarizations using equations (36) and were rotated to obtain the maximum and minimum eigen-impedances using the ASVD formulation. The resistivity values and contrasts between the structural elements of the model were selected so as to generate long delays and phases close to 90°, on the conductive side of the interfaces; responses with marginal phases would, in theory, be easier to destabilize with distortion and/or noise. Thus, the "measurements' were conducted on the conductive side of the interface, above the north-eastern corner of the tube at the location indicated by the downward pointing black triangle: this will henceforth be referred to as *Site X*.

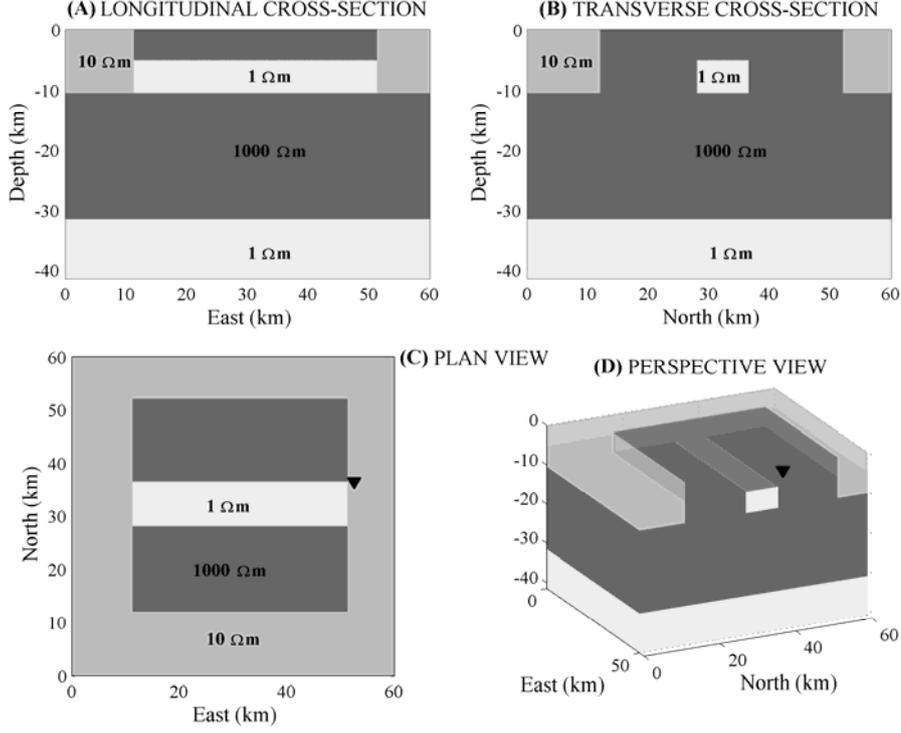

**Figure 1.** The 3-D model used to generate the synthetic data used in Section 4.1. The downward pointing black triangle indicates the location of *Site X*, where the "measurements" yielding the tensor analysed in Figures 2, 3 and 4 were obtained.

Fig. 2a shows the apparent resistivities and phases of the tensor elements at *Site X*. The phases of the off-diagonal elements remain strictly within quadrant ([0°, 90°] for $Z_{xy}$ and [180°, 270°] for $Z_{yx}$, as expected. Moreover, they are marginal and approximate the 90°/270° upper limit at certain period ranges, with particular reference to the phase of $Z_{yx}$. The phases of the diagonal elements are clearly seen to move in and out of the first and third quadrants, particularly at periods shorter than 20s – 30s, which roughly correspond to the inhomogeneous upper layer and the transition from the upper to the middle layer. Fig. 2b illustrates the almost exact induced (output) electric fields used in computing the tensor. Quite clearly, the components $E_x^{(2)}$ and $E_y^{(1)}$ are responsible for the anomalous phases observed in the diagonal elements $Z_{xx}$ and $Z_{yy}$. In particular, $E_x^{(2)}$ is weak in the period interval 1s – 20s, where it produces the bizarre phase variation observed at the same band of $Z_{xx}$. $E_y^{(1)}$ *changes orientation* from 'normal' (third quadrant) at $T < 0.2$s to 'reverse' (first quadrant) at $T > 10$s and, in turn, causes the phase of $Z_{yy}$ to wrap around the unit circle.

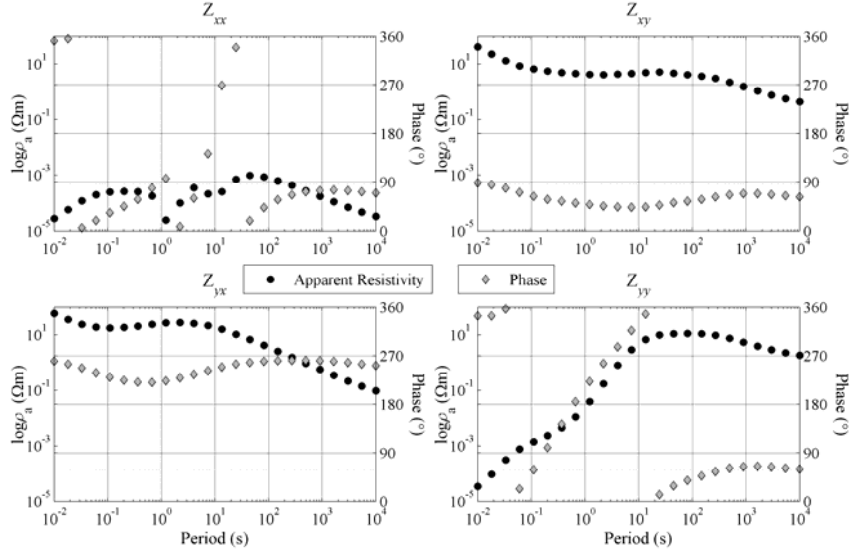

**Figure 2a.** The almost exact impedance tensor "observed" at *Site X* of the 3-D model shown in Fig. 1, when the horizontal electric and magnetic fields are measured in directions parallel and perpendicular to the horizontal coordinate axes (model axes).

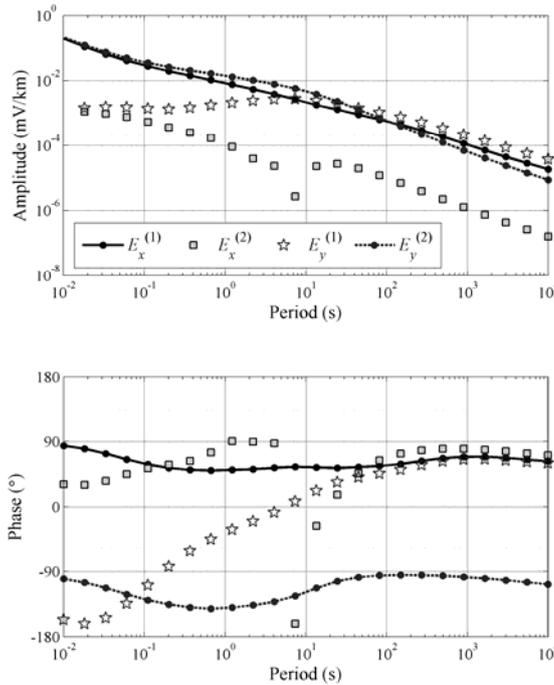

**Figure 2b.** The induced electric field components observed at *Site X* of the 3-D model shown in Fig. 1. The components $E_x^{(1)}$ and $E_y^{(1)}$ correspond to the fields induced by a unit amplitude magnetic field polarized parallel to the *x*-axis of the model (Polarization 1). The components $E_x^{(2)}$ and $E_y^{(2)}$ correspond to the fields induced by a unit amplitude magnetic field polarized parallel to the *y*-axis (Polarization 2).

Note, however, that while $E_x^{(2)}$ clearly exhibits a short-lived excursion from its stable passive state in the first quadrant, $E_y^{(1)}$ shifts from a stable 'normally' oriented passive state at short periods to a stable reversely oriented passive state at long periods, in a rather orderly manner. In both case, the transitions involve the *fourth* quadrant, so that the real parts of $E_x^{(2)}$ and $E_y^{(1)}$ remain strictly positive at all times and their passivity is never challenged! Moreover, because the components $E_x^{(1)}$ and $E_y^{(2)}$ are strictly passive at all periods, the tensor is dominated by the stable, causal and passive off-diagonal elements.

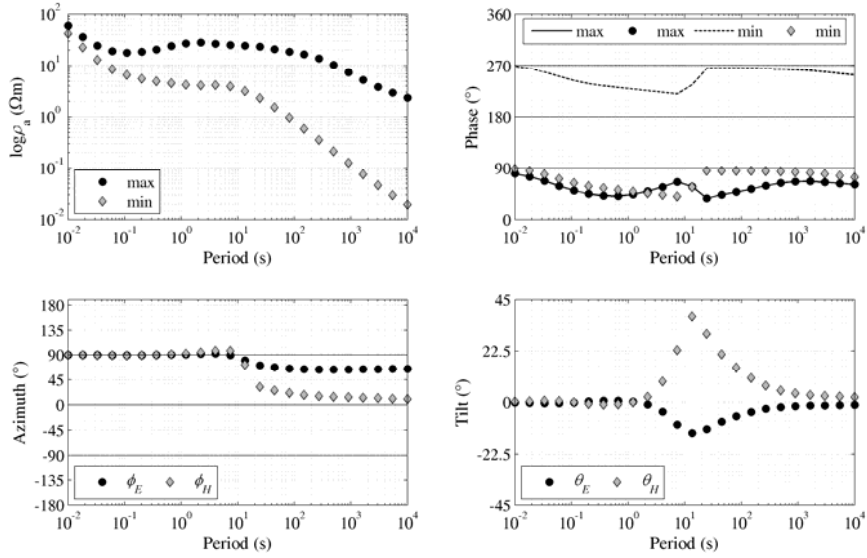

**Figure 2c.** The characteristic states of the impedance tensor observed at *Site X* of the 3-D model shown in Fig. 1. The top-left panel shows the maximum and minimum apparent resistivities corresponding to the maximum and minimum eigen-impedances respectively. The top-right panel illustrates the maximum and minimum phases corresponding to the maximum and minimum eigen-impedances respectively. The phases are computed with both the two-quadrant arc-tangent function (discrete symbols) and the four-quadrant arc-tangent function (lines); the double mode of display is justified in the text. The bottom-left panel illustrates the azimuths $\varphi_E$ and $\varphi_H$ of the electric and magnetic eigen-frames respectively. The middle-right panel shows the tilt angles $\theta_E$ and $\theta_H$ of the electric and magnetic eigen-frames respectively.

Fig. 2c shows the characteristic states of the impedance tensor at *Site X*. The top-right panel illustrates the phases of the eigen-impedances which are plotted twice: once in the form of discrete lines, computed with the two-quadrant arc-tangent function, and once in the form of continuous lines, computed with the four-quadrant arc-tangent function. The double mode of display is adopted in order (a) to evaluate the passivity of the eigen-impedances by reference to the definition (containment) of their phases in the first quadrant, (b) to evaluate the passivity of the eigen-impedances by reference to the equivalence or oppositeness of the signs of their real and imaginary parts, and, (c) to observe and evaluate reversals of the signs of the eigen-impedances signifying reversals (inversions) of the coordinate axes. It is apparent that the real and imaginary parts of the maximum eigen-impedance are both positive so that the maximum phase is confined in first quadrant. As also expected, the real and imaginary parts of the minimum eigen-impedance are both negative and the minimum phase is contained in the third or first quadrant, depending on the arc-tangent function used. Both eigen-impedances behave as required of passive functions defined in the practical Magnetotelluric coordinate system.

As mentioned above, the passivity of the Earth response at *Site X* has never actually been challenged and is further assured by dominant contributions of $E_x^{(1)}$ and $E_y^{(2)}$ at all periods. As it turns out, the anomalous phases of the diagonal elements are only *apparent* or, so to speak, epiphenomenal effects of the mixing of phases during impedance tensor calculations in a particular coordinate system: the tensor of Fig. 2a was calculated in an "experimental" frame that coincides with the principal directions of the geoelectric structure ((intrinsic coordinate frame). As it is rather straightforward to verify, if the tensor was "measured" in an "experimental" frame rotated to an angle > 40° with respect to the intrinsic, the phases of all four elements would be bounded in [0°, 90°].

An additional significant observation is that at periods 5 – 50 s, where $Z_{xx}$ and $Z_{yy}$ exhibit their strongest anomalous phase variation, the phases of both eigen-impedances exhibit a rather sharp inflection (Fig. 2c, top-right). This is matched by corresponding variations in the geometry of the electric and magnetic eigen-frames (Fig. 2c, bottom-left). The azimuths $\varphi_E$ and $\varphi_H$ of the electric and magnetic eigen-frames are approx. 90° up to $T \approx 10s$, inasmuch as the "measurements" were conducted parallel and perpendicular to the horizontal axes of geoelectric (intrinsic) coordinate system; the azimuths deviate to the north at $T > 10s$. The bottom-right panel illustrates the tilts $\theta_E$ and $\theta_H$ of the respective eigen-frames. These very small (the eigen-frames are practically horizontal) up to $T \approx 2s$ but explode thereafter and maximize at $T \approx 10s$, indicating a steep local eigen-field topography and high gradients; accordingly, the polarization of the total magnetic field goes almost circular and the polarization of the electric field significant. The magnitude of the tilts diminishes afterwards and returns to horizontal (smooth field topography) at long periods. The commensurate behaviour of the eigen-impedance phases and eigen-frame geometries correspond to the period interval at which the field transits the depth range of the conductive tube and signify the strong influenced of its presence.

The tensor of *Site X* is "normal" in the sense that it only exhibits diagonal anomalous phases which are generally associated with tributary (and in this case passive) coupling modes that do not affect its passive nature. It would be interesting, however, to try changing its constitution in ways purportedly producing violation of passivity, as for instance by reversing the orientation of the electric field. In order to cover all possibilities, this exercise was performed by subjecting the tensor to transitional reversal (inversion) of the experimental electric coordinate frame by rotation. Thus, in the distortion model of eq. (37), $\phi(\omega)$ is allowed to vary in the interval [0, $\pi$], and the remaining parameters are fixed at $a = 0$ so that $\mathbf{A} = \mathbf{I}$, $\xi = 0$ so that $\mathbf{B} = \mathbf{I}$ and $\hat{\mathbf{e}}^{(k)}(\omega) = \mathbf{0}$. This operation will produce complete inversion of one coordinate axis when $\phi(\omega) \geq \pi/2$ and complete inversion of both axes when $\phi(\omega) \geq \pi$.

Letting $\mathcal{Z}[\phi(\omega)]$ denote the distorted (in this case twisted) impedance tensor, the apparent resistivities and phases of the distorted tensor elements produced by this experiment are shown in Fig 3a. They are definitely strange entities with peculiar apparent resistivity curves and phases that typically vary in the first and third quadrants with the exception of non-passive (second quadrant) transitional anomalous phases observed in $\mathcal{Z}_{xy}[\phi(\omega)]$ for $T \in (4s, 20s)$ and in $\mathcal{Z}_{yy}[\phi(\omega)]$ for $T \in (60s, 140s)$. The discontinuity in the phase of $\mathcal{Z}_{yx}[\phi(\omega)]$ corresponds to the point at which $\phi(\omega)$ straddles the $\pi/2$ boundary.

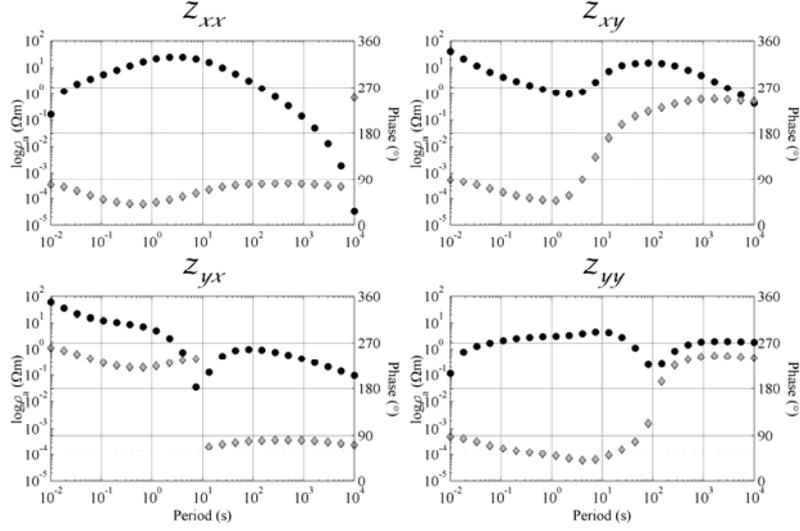

**Figure 3a.** The impedance tensor obtained after subjecting the almost exact electric fields observed at *Site X* (Fig. 2b) to transitional reversal of the experimental electric coordinate frame through isometric transformation (rotation) on the horizontal plane by an angle linearly varying in the interval $[0, \pi]$. The layout of the panels is as per Fig. 2a.

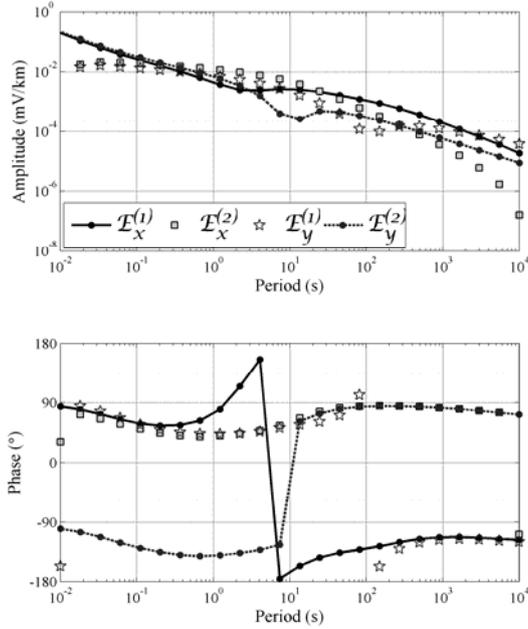

**Figure 3b.** The induced electric field components observed at *Site X* (Fig. 2b) under transitional reversal of the experimental electric coordinate frame by rotation through an angle linearly varying in the interval $[0, \pi]$. $\mathcal{E}_x^{(1)}$ and $\mathcal{E}_y^{(1)}$ correspond to the rotating field components of Polarization 1; $\mathcal{E}_x^{(2)}$ and $\mathcal{E}_y^{(2)}$ correspond to the rotating field components of Polarization 2.

The electric fields producing the tensor of Fig. 3a are shown in Fig. 3b. It is apparent that three components of the electric field reverse orientation as evident in their phases changing quadrants from first to third for $\mathcal{E}_x^{(1)}[\phi(\omega)]$ and $\mathcal{E}_y^{(1)}[\phi(\omega)]$ and from third to first for $\mathcal{E}_y^{(2)}[\phi(\omega)]$. The phases of the former two components flip when $\phi(\omega)$ crosses $\pi/2$ and the phase of the last when it crosses $3\pi/4$. Moreover, the transition of $\mathcal{E}_x^{(1)}[\phi(\omega)]$, $\mathcal{E}_y^{(1)}[\phi(\omega)]$ between the first and third quadrants takes place through the second quadrant, in which they acquire a negative real part (apparent violation of passivity). It is evident, however, that during the transition the amplitudes of these components are considerably weaker than the corresponding amplitudes of the strictly passive $\mathcal{E}_x^{(2)}[\phi(\omega)]$ and $\mathcal{E}_y^{(2)}[\phi(\omega)]$. In consequence, the parallel filter rule will ensure that the linear superposition of the passive and (apparent) non-passive processes will be dominated by the passive processes and will not affect the overall constitution of the tensor, as will be amply demonstrated forthwith.

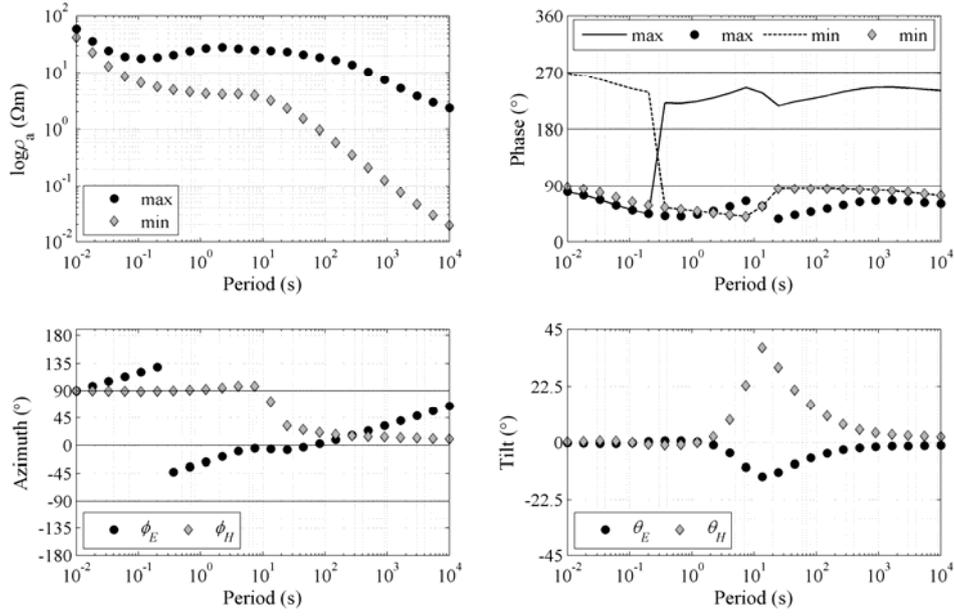

**Figure 3c.** The characteristic states of the impedance tensor computed from the components $\mathcal{E}_x^{(1)}$, $\mathcal{E}_y^{(1)}$ $\mathcal{E}_x^{(2)}$ and $\mathcal{E}_y^{(2)}$ of Fig 3b, obtained by subjecting the almost exact electric fields observed at *Site X* (Fig. 2b) to transitional reversal by rotation in the interval [0, $\pi$]. The layout of the panels and mode of presentation is exactly as per Fig. 2c.

Fig. 3c illustrates the characteristic states of the rotated tensor. It is immediately evident that the maximum and minimum apparent resistivities and phases are recovered without loss (Fig. 3c, top-left and top-right): they are exactly the same with the corresponding parameters observed in Fig. 2c. The real and imaginary parts of both eigen-impedances always carry the same sign; they do flip sign upon inversion of the coordinate frame but they do so in unison and without changing their passive character (Fig. 3c, top-right). The continuous rotation is also evident in the variation of the electric eigen-frame's azimuth ($\varphi_E$, Fig. 3c, bottom-left) while the electric eigen-frame's tilt ($\theta_E$) is not affected by the rotation on the horizontal plane and the magnetic eigen-frame's geometry is not affected at all because nothing was done to the magnetic field's experimental coordinate frame.

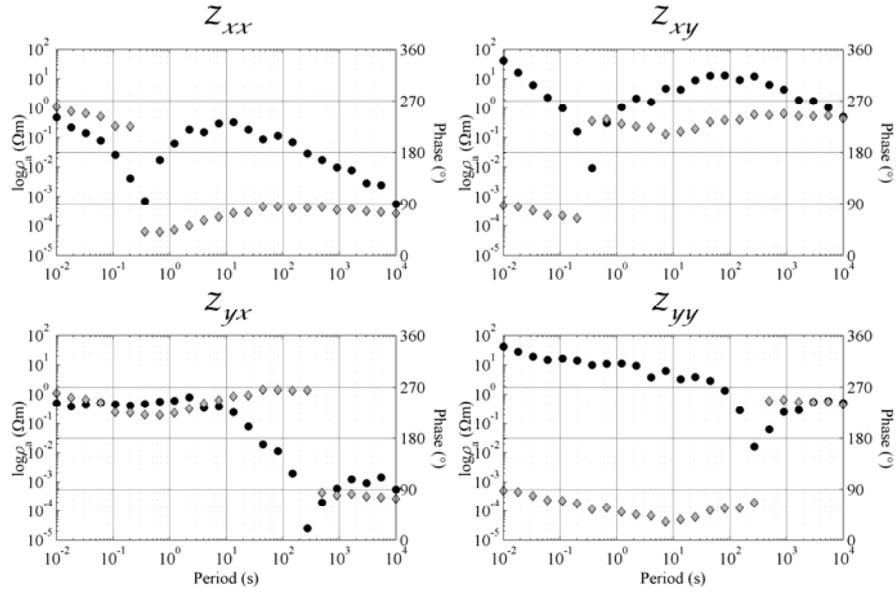

**Figure 4a.** The impedance tensor obtained after subjecting the almost exact electric fields observed at *Site X* (Fig. 2b) to shifting (splitting), shearing and transitional reversal of the experimental electric coordinate frame by rotation in the interval [0, $\pi$]. The layout of the panels is as per Fig. 2a.

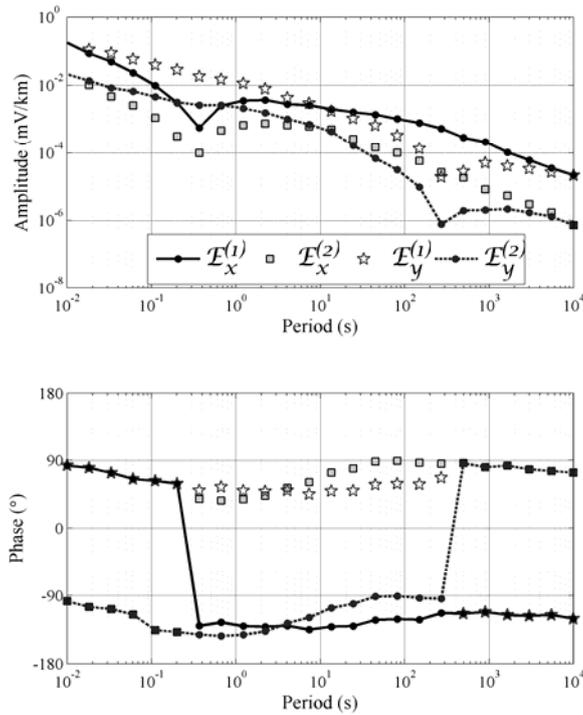

**Figure 4b.** The induced electric field components observed at *Site X* (Fig. 2b) after splitting, shearing and transitional reversal by rotation in the interval [0, $\pi$]. $\mathcal{E}_x^{(1)}$ and $\mathcal{E}_y^{(1)}$ correspond to the distorted components of Polarization 1; $\mathcal{E}_x^{(2)}$ and $\mathcal{E}_y^{(2)}$ to the distorted components of Polarization 2.

A third and final experiment investigates the behaviour of the data of *Site X* under transitional coordinate frame inversion of a *sheared and shifted* electric field, as specified by the distortion model of eq. (37). Thus, $\phi(\omega)$ is again allowed to vary in the interval [0, $\pi$], but this time $a = 0.8$, $\xi = 44.8°$, and 10% random noise has been added to the electric and magnetic field components. It is probably useful to emphasize the maximum possible shearing is effected at $\xi = 45°$ and the case $\xi = 44.8°$ used herein is very extreme; for details see Groom and Bailey (1989).

Fig. 4a illustrates the elements of the distorted tensor, in the form of apparent resistivities and phases. The effect of shearing is evident in that $Z_{xy}[\phi(\omega)]$ and $Z_{yy}[\phi(\omega)]$ have now become the dominant elements of the tensor and $Z_{xx}[\phi(\omega)]$, $Z_{yx}[\phi(\omega)]$ the tributary. The phases of the dominant elements flip in response to the rotating coordinate frame, but only with discontinuous jumps from one passive state to another: they are always strictly contained in the first and third quadrants. The phases of the dominant and tributary tensor elements are also stable and confined in the first and third quadrants. Fig. 4b shows the distorted electric field components. As should have been expected from the study of Fig. 4a, the shearing has conferred dominance to $\mathcal{E}_x^{(1)}[\phi(\omega)]$ and $\mathcal{E}_y^{(1)}[\phi(\omega)]$, whose phases flip between reversed passive states in the first and third quadrant ($\mathcal{E}_x^{(1)}[\phi(\omega)]$) or are stably confined in the third quadrant ($\mathcal{E}_y^{(1)}[\phi(\omega)]$). The same but exactly symmetric behaviour is observed in the tributary distorted components, $\mathcal{E}_x^{(2)}[\phi(\omega)]$ and $\mathcal{E}_y^{(2)}[\phi(\omega)]$.

Fig. 4c shows the corresponding characteristic states. The maximum and minimum apparent resistivities are shifted apart by approx. six orders of magnitude and the information they originally contained is practically lost (top-left). The maximum and minimum phases fare better (top-right) but the im-portant information about the conductive tube, originally contained in the interval 5 – 50s is also lost (cf. top-right panel of Fig. 2c). Still, they are both contained in the first quadrant when computed with the two-quadrant arc-tangent function. The real and imaginary parts of the maximum eigen-impedance $\zeta_1$ are both positive for $T < 10$s and both negative for $T > 10$s; they flip in unison as $\phi(\omega)$ crosses the $\pi/2$ boundary. The same but opposite behaviour is observed in the real and imaginary parts of the minimum eigen-impedance $\zeta_2$. Thus, in spite of the distortion, both eigen-impedances behave as causal and passive functions. The spatial information originally contained in the electric and magnetic eigen-frames is also completely lost. Only the electric "eigen-azimuth" $\varphi_E$ exhibits variation in response to the rotation of the sheared electric field (bottom-left): it starts at exactly $\varphi_E = 44.8°$, which is the direction of shearing of the un-rotated (untwisted) tensor at $T = 0.01$s and varies linearly to $\phi(T \cong 7s) = 131°$; it is automatically reset to $\phi(T \cong 13.5s) = -41.3°$ when $\phi$ crosses the $\pi/2$ boundary and the $y$-axis reverses orientation, thereafter varying linearly to $\varphi_E = 44.8°$ when $\phi(T = 10,000s) = 180°$.

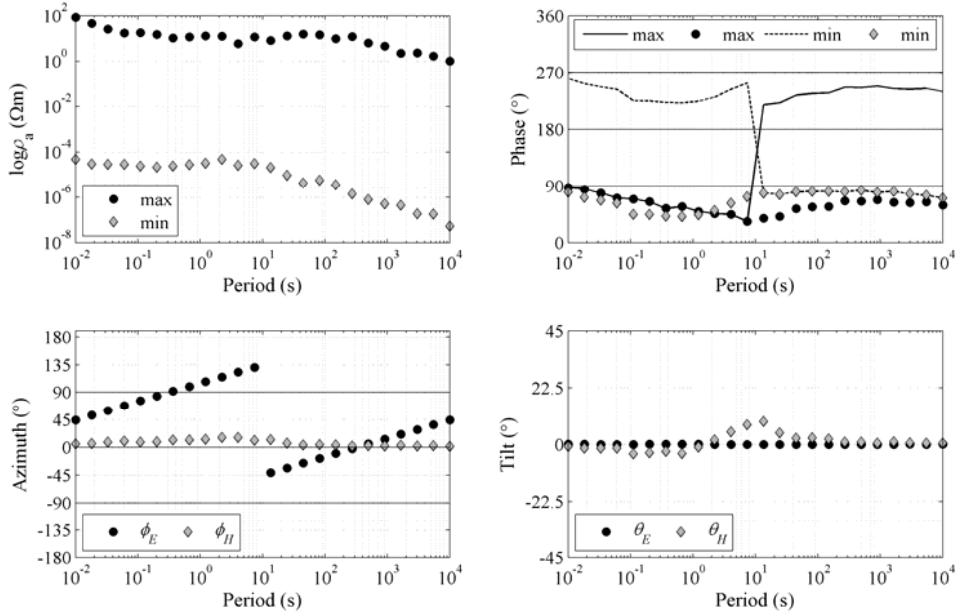

**Figure 4c.** The characteristic states of the impedance tensor computed from the components $\mathcal{E}_x^{(1)}$, $\mathcal{E}_y^{(1)}$ $\mathcal{E}_x^{(2)}$ and $\mathcal{E}_y^{(2)}$ of Fig 4b, obtained by subjecting the almost exact electric fields observed at *Site X* (Fig. 2b) to splitting, shearing and transitional reversal by rotation in the interval [0, $\pi$]. The layout of the panels and mode of presentation is exactly as per Fig. 2c.

The results presented in Section 4.1 can be briefly summarized as follows: When the dominant (usually off-diagonal) coupling modes are causal and passive, electric field reversals may produce transitional anomalous phases but do not violate the passivity of the impedance tensor and its eigen-impedances. Non-passive tributary coupling modes that may appear as a result of the response to different source polarizations, or coordinate frame rotations, are generally absorbed by the dominant passive modes so that the total electric and magnetic fields remain passive. Likewise, when the dominant coupling modes are passive, distortion by real (non-reactive) operators may change the amplitudes of the electric fields, superimpose electric field components and mix their phases, as well as annihilate Earth structural information; however, it does not in itself possess phase-shifting capacity so as alter the time-history of the dominant modes and threaten their passivity. Although based on a single example, these conclusions are articulated in an affirmative way because they are founded on solid theoretical ground (the parallel filter rule and systems theory) and are therefore valid under all similar circumstances.

**4.2. Anisotropic 2-D impedance tensor with diagonal and off-diagonal anomalous phases.**

If the tensor of *Site X* is "more or less normal", the next example is concerned with a case that is genuinely not. Synthetic 2-D impedance tensors with large anomalous phase variations in both diagonal and off-diagonal elements have been presented by Heise and Pous (2003) on the basis of a model comprising an azimuthally anisotropic block resting on an azimuthally anisotropic layer, with the anisotropy strikes of the block and the layer being orthogonal. The model and some responses are clearly illustrated in Fig. 1 of their paper; the model is also described in detail sufficient to warrant a faithful as possible reproduction. The reproduction is not shown herein for the sake of conciseness. As in Heise and Pous (2003), the model was solved with the two-dimensional finite difference anisotropic forward algorithm of Pek and Verner (1997). The response used herein was obtained at a location at the surface of the model corresponding to the location of *Site 4* shown in Fig. 1 of Heise and Pous.

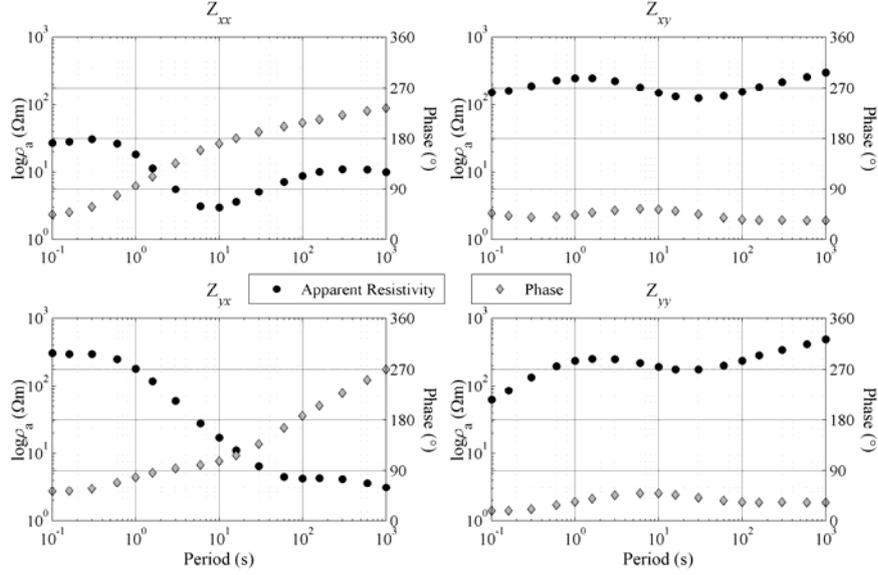

**Figure 5a.** The noise-free almost exact impedance tensor "observed" at *Site 4* of the 2-D anisotropic model shown in Figure 1 of Heise and Pous (2003), when the horizontal electric and magnetic fields are measured in directions parallel and perpendicular to the regional 2-D strike. The layout of the panels is as per Fig. 2a.

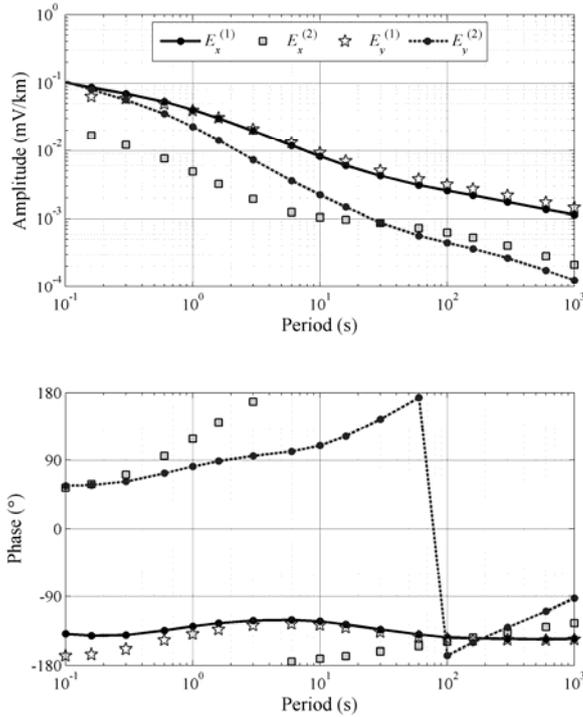

**Figure 5b.** The induced electric field components observed at *Site 4* of the 2-D anisotropic model shown in Figure 1 of Heise and Pous (2003). The components $E_x^{(1)}$ and $E_y^{(1)}$ correspond to the fields induced by a unit amplitude magnetic field polarized parallel to the regional 2-D strike (x-axis of the model – Polarization 1). The components $E_x^{(2)}$ and $E_y^{(2)}$ correspond to the fields induced by a unit amplitude magnetic field polarized perpendicular to the regional 2-D strike (Polarization 2).

The synthetic response of *Site 4* is shown in Fig. 5a, in the form of apparent resistivities and phases of the tensor's elements. The phases of $Z_{xx}$ and $Z_{yx}$ are clearly seen to rotate continuously from the first via the second to the third quadrant, exhibiting negative real parts during their transition through the se-cond quadrant in the period intervals 1 – 10s and 1 – 80s respectively. The phases of $Z_{xy}$ and $Z_{yy}$ are stable and confined in the first quadrant. Heise and Pous (2003) demonstrated that the interaction between the anisotropic block and anisotropic layer produces *complete* reversal in the orientation of the $E_y$ component. They attributed the anomalous phases, at least in part, to this reversal. It is hard to determine whether the passive right column elements ($Z_{xy}$ and $Z_{yy}$) are dominant at short periods, because, as can be inferred by inspecting apparent resistivities, they have amplitudes comparable to $Z_{yx}$ while the amplitude of $Z_{xx}$ is also significant. They appear, however, to become dominant at long periods where the amplitudes of the left column elements attenuate rapidly and significantly. Such questions can be answered by studying the electric fields observed at the location of *Site 4*, which are shown in Fig. 5b. It can easily be seen that at short periods, the components $E_x^{(1)}$, $E_y^{(1)}$ and $E_y^{(2)}$ are significant and comparable of amplitude. $E_x^{(2)}$ is tributary and approx. one order of magnitude weaker than the other components; moreover, it begins to reverse orientation almost immediately, moving into the second quadrant at $T > 0.6$s and stabilizing in the third quadrant after $T = 6$s. These observations may adequately explain the low amplitude and anomalous phase of $Z_{xx}$. The $E_y^{(2)}$ component begins to attenuate after $T \approx 1$s at a rate much faster than the rate experienced by the other three components, and at the same time to reverse orientation by migrating into the first quadrant via the *second* (non-passive behaviour). By $T \approx 100$s its amplitude is comparable but less than that of $E_x^{(2)}$ and its orientation has totally flipped and stabilized in the first quadrant. These properties also suffice to explain the diminishing amplitude and anomalous phase of $Z_{yx}$. The $E_x^{(1)}$ and $E_y^{(1)}$ components have *comparable*, almost equal amplitudes and their phases are almost identical and firmly defined in the first quadrant. In addition, they are significantly to almost one order of magnitude stronger than $E_x^{(2)}$ and $E_y^{(2)}$ at periods longer than approx. 1s. This also explains the stability of $Z_{xy}$ and $Z_{yy}$, which out to comprise the dominant passive elements of a tensor almost exclusively defined by the response of polarization 1!

The characteristic states of the tensor are shown in Fig. 5c. The phases of the maximum and minimum eigen-impedances are both defined in the first quadrant and, as could have been anticipated, the real and imaginary parts of both eigen-impedances carry the same sign (top-left). It is therefore apparent that the eigen-impedances and, in consequence, the tensor, are causal and passive functions. It is also worth noting that the real and imaginary parts of the minimum eigen-impedance are positive, as a consequence of both $E_x^{(1)}$ and $E_y^{(1)}$ being defined in the first quadrant due to the complete reversal of the total $E_y$ component (for details see Heise and Pous, 2003). The azimuth $\varphi_E$ of the electric eigen-frame (bottom-left) is practically fixed at 49°–51° except for the shortest periods. Likewise, the tilt $\theta_E$ of the electric eigen-frame is small but finite up to $T \approx 10$s and disappears afterwards (bottom-right).

At periods longer that approx. 3s, the near equality of the dominant $E_x^{(1)}$ and $E_y^{(1)}$ components indicates that the electric field at *Site 4* is *naturally sheared* to an almost *maximal* degree as a result of the peculiar conductivity structure (orthogonality of the anisotropy strikes). This would also explain the near-invariance and value of the electric eigen-frame's azimuth at the same period range (and would also compare to the behaviour of the artificially sheared and twisted electric fields at *Site X*, illustrated in Fig. 4b. As it turns out, this particular conductivity configuration suppresses the electric fields of polarization (2) and allows the electric fields of polarization (1) to dominate the induction process, but only after deforming and adjusting them to the only locally permissible propagation path into the Earth. The suppression of polarization (2) involves a rather intense reactive effect whose clear manifestation is the high attenuation rate of experienced by $E_y^{(2)}$ in the interval 1s – 100s, as well as the significant delay (phase shift) that pushes its phase into the second quadrant and destroys its passivity. The exact nature of the effect will not be presented herein because an analogous (and more instructive) 3-D phenomenon will be scrutinized in Section 4.3 below. Nevertheless, in support of this interpretation, Fig. 5d illustrates the characteristic states of a *distorted* impedance tensor, produced by *additionally* shearing the electric fields using the parameters $a = 0$, $\phi(\omega) = 0\ \forall\ \omega$, $\xi = 44.8°$ and $\hat{\mathbf{e}}^{(k)}(\omega) = \mathbf{0}$. The effects of the additional extreme shearing are: (i) Static shift of the maximum and minimum apparent resistivities without, as can easily be verified, significant distortion of their shape; (ii) firm fixing of the electric eigen-frame's azimuth at 45°; (iii) elimination of the electric-eigen-frame's tilt. Not surprisingly, there are practically no effects on the phases of the eigen-impedances and the magnetic eigen-frame's geometry. This means that the additional shearing has eliminated the tributary contribution of polarization (2) but, with the exception of the static shift, has left the dominant passive contribution of polarization (1) practically intact.

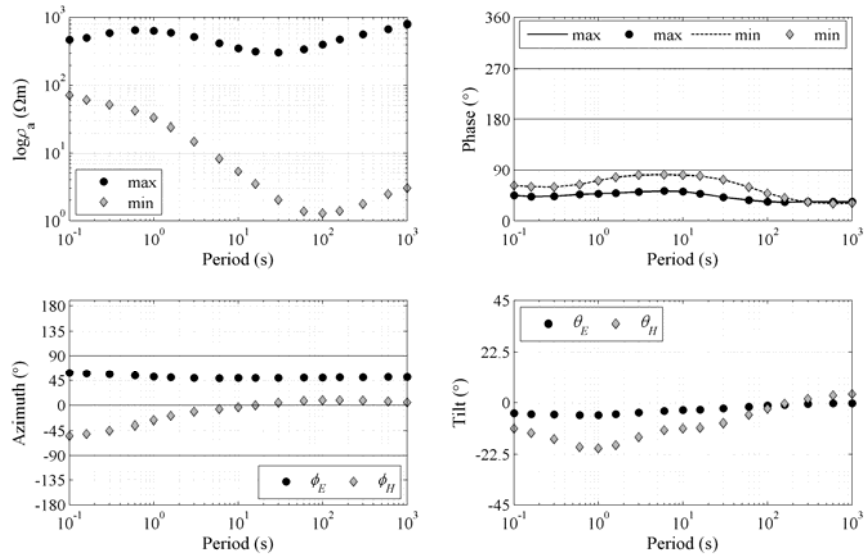

**Figure 5c.** The characteristic states of the almost exact impedance tensor at *Site 4* of the 2-D anisotropic model shown in Figure 1 of Heise and Pous (2003). The layout of the panels and mode of presentation is exactly as per Fig. 2c.

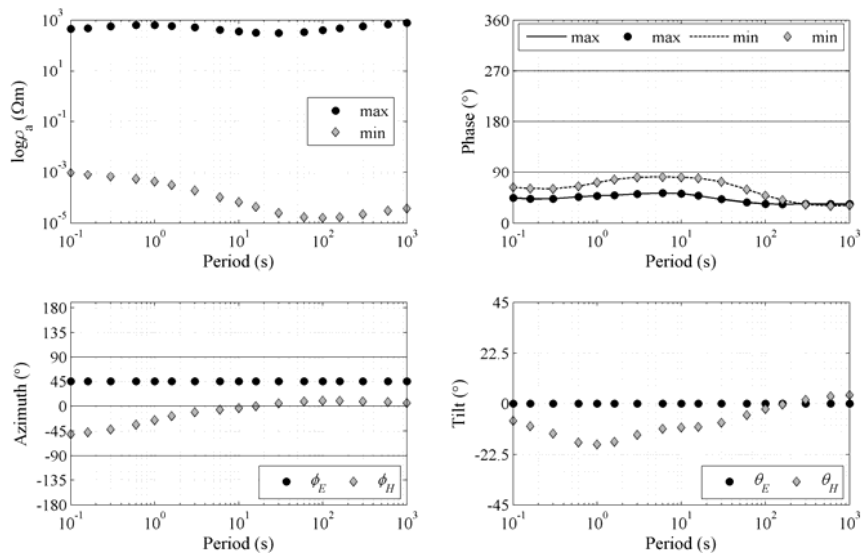

**Figure 5d.** The characteristic states of a distorted tensor, obtained after subjecting the almost exact electric fields observed at *Site 4* (Fig. 5b) to additional extreme shearing by an angle $\xi = 44.8°$. The layout of the panels and mode of presentation is exactly as per Fig. 2c.

To conclude this example, note that whether the natural shearing of the electric field qualifies as "distortion" of the impedance tensor or not is an altogether different question that has more to do with the definition of distortion. At any rate, this tensor conveys valuable information about the geoelectric structure and can be used for interpretation, as was plainly demonstrated by Heise and Pous (2003). Moreover, it is apparent that the naturally reversed and sheared electric fields may generate anomalous phases, but if the linearity of the tensor is not challenged its passivity is also not challenged and the anomalous phases are inconsequential.

**4.3. 3-D impedance tensor with diagonal and off-diagonal anomalous phases.**

The model used to generate the 3-D response is shown in Fig. 6. It comprises a 15km thick upper inhomogeneous layer, overlying a homogeneous resistive (1000 Ωm) half-space. The inhomogeneous upper layer consists of a 'regional' (40×30×15 km) parallelepiped block with sea-water conductivity (0.3 Ωm), embedded in a 1000 Ωm resistive host. A 5×15×15 km 'local' conductive (3 Ωm) parallelepiped block is also embedded in the resistive host at the northeast corner of the sea-water conductor, so that together they form an L-shaped conductive inhomogeneity. The induced fields were computed with the finite difference algorithm of Mackie and Madden, (1993) and Mackie Madden and Wannamaker, (1993). The Earth responses were computed on the basis of field components generated by two orthogonal source polarizations using equations (36) and were rotated to obtain the maximum and minimum eigen-impedances using the ASVD formulation. The model is simple, but the configuration of the conductors and the conductivity contrasts were chosen so as to generate large field gradients and, hopefully, in the bay formed by the regional and local conductors, scattering that would create a secondary effect equivalent to an internal source. Thus, the "measurements' were conducted at the resistive side of the western interface between the host and the local conductor, but marginally close to the interface, at the location indicated by the downward pointing black triangle: this will henceforth be referred to as *Site Y*.

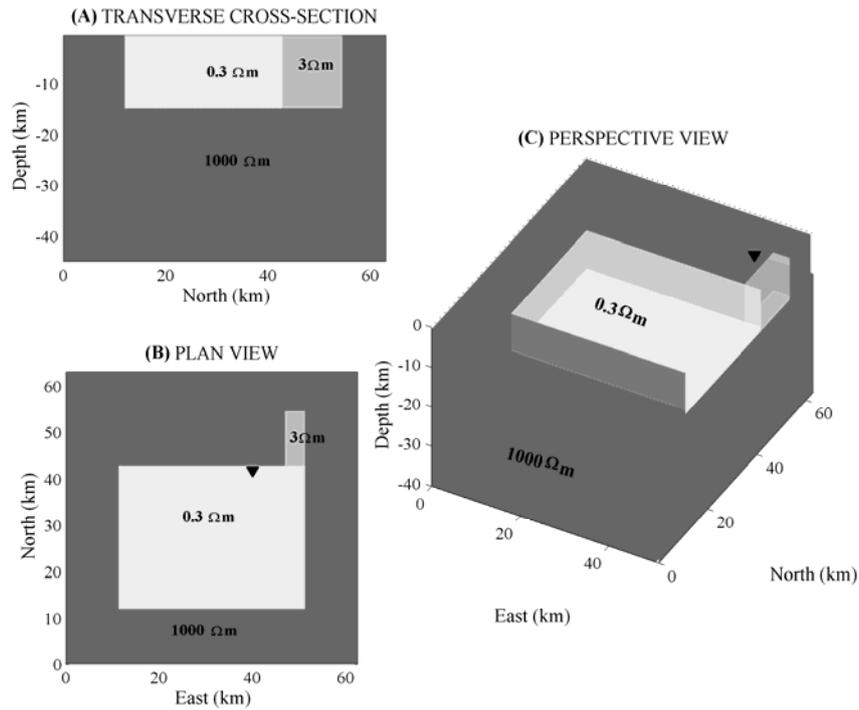

**Figure 6.** The 3-D model used to generate the synthetic data used in Section 4.3. The downward pointing black triangle indicates the location of *Site Y*, at which the "measurements" yielding the tensor analysed in Figures 7a – 7c were obtained.

The synthetic response at *Site Y* is shown in Fig. 7a. It is immediately apparent that the elements $Z_{xx}$ and $Z_{xy}$ vary in an opposite, although not symmetric manner. At the outset, the apparent resistivity of $Z_{xx}$ is very low and its phase is located in third quadrant. However, the apparent resistivity increases rapidly by approx. five orders of magnitude and the phase rotates via the fourth to the first quadrant, until at periods $T > 1$s the apparent resistivity it asymptotically approaches 1500Ωm and the phase 45°. Evidently, the real part of $Z_{xx}$ is positive throughout. On the other hand, the apparent resistivity of $Z_{xy}$ begins at approx. 1700 Ωm and its phase at 45°; the apparent resistivity gradually declines by approx. two orders of magnitude and its phase rotates through the second to the third quadrant in the interval 15s $< T <$ 500s, during which it acquires a negative real part (non-passive state). Of the other two elements, $Z_{yx}$ exhibits a rather 'normal' behaviour and its phase is firmly defined in the third quadrant. On the other hand, the apparent resistivity of $Z_{yy}$ is generally low ($< 1$Ωm for $T < 40$s and less than 3Ωm overall) and its phase describes an almost complete rotation around the unit circle for $T < 40$s, before it stabilizes in the first quadrant. In the interval 0.1s $< T <$ 0.4s the phase transits the second quadrant and $Z_{yy}$ has a negative real part.

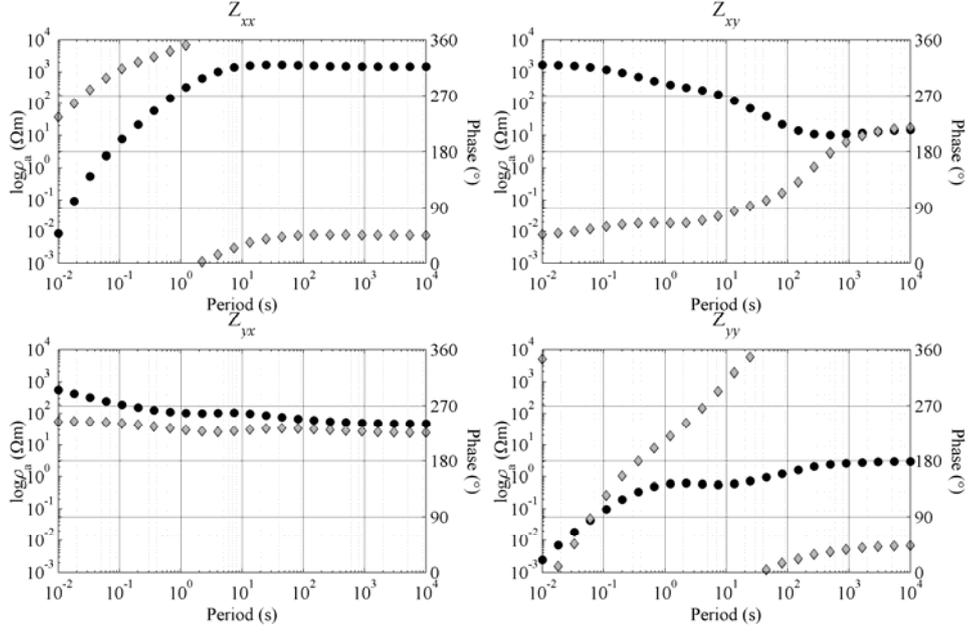

**Figure 7a.** The almost exact impedance tensor "observed" at *Site Y* of the 3-D model shown in Fig. 6, when the horizontal electric and magnetic fields are measured in directions parallel and perpendicular to the horizontal coordinate axes (model axes). The layout of the panels is as per Fig. 2a.

Explanations of this behaviour will be first sought by studying the induced electric fields, which are shown in Fig. 7b. At very short periods the conditions approximate a half-space in which $E_x^{(1)}$ and $E_y^{(2)}$ are dominant and defined in the first and third quadrants respectively; they are almost two orders of magnitude stronger than the tributary $E_x^{(2)}$ and $E_y^{(1)}$ which are both defined in the fourth quadrant and have a positive real part. The higher structural dimensions almost immediately begin to manifest their presence. $E_x^{(2)}$ *gains* strength in the interval 0.05s – 10s and its phase rotates from the fourth to the first quadrant, crossing the boundary at the period $T \approx 3$s where it also attains its maximum amplitude; in the interval 5s – 100s it becomes dominant and begins to attenuate exponentially, with its phase fluctuating in the first quadrant until it stabilizes at approx. 45°. The phase of $E_x^{(1)}$ begins to rotate into the second quadrant after approx. 0.1s, crossing the boundary at $T \approx 3$s, almost simultaneously with $E_x^{(2)}$ crossing into the first quadrant. In the interval 3s – 100s $E_x^{(1)}$ experiences power-law attenuation at a rate considerably higher than that of the other three components, while its phase continues rotating counter-clockwise until, at approx. 80s, crosses into the third quadrant; thereafter, $E_x^{(1)}$ attenuates exponentially and completely reverses orientation after approx. 200s, with its phase asymptotically approaching -135°. $E_y^{(1)}$ remains weak (tributary) throughout, but does not lose power up to approx 0.8s, thereafter attenuating exponentially; its phase rotates almost immediately (at $T > 0.05$s) from the fourth into the first quadrant where is fluctuates until is asymptotically approaches 45° after $T \approx 100$s. Finally, $E_y^{(2)}$ decays exponentially, with its phase firmly defined in the third quadrant. According to the above observations, the variation of $E_x^{(1)}$ in the interval 1s – 100s clearly indicates the existence of a reactive effect that attenuates the component and introduces delay that destroys its passivity. Moreover, the reversal the same component through the non-passive domain (second quadrant) would appear to explain the corresponding non-passive constitution of $Z_{xy}$ at the approx. same period interval. Similarly, the variation of $E_x^{(2)}$ is apparently manifest in $Z_{xx}$ and the variation of $E_y^{(2)}$ in $Z_{yx}$. Finally, the bizarre phase variation of $Z_{yy}$ at $T < 10$s is evidently a result of the weakness of $E_y^{(1)}$ and the consequent mixing of phases with $E_y^{(2)}$ and the magnetic field.

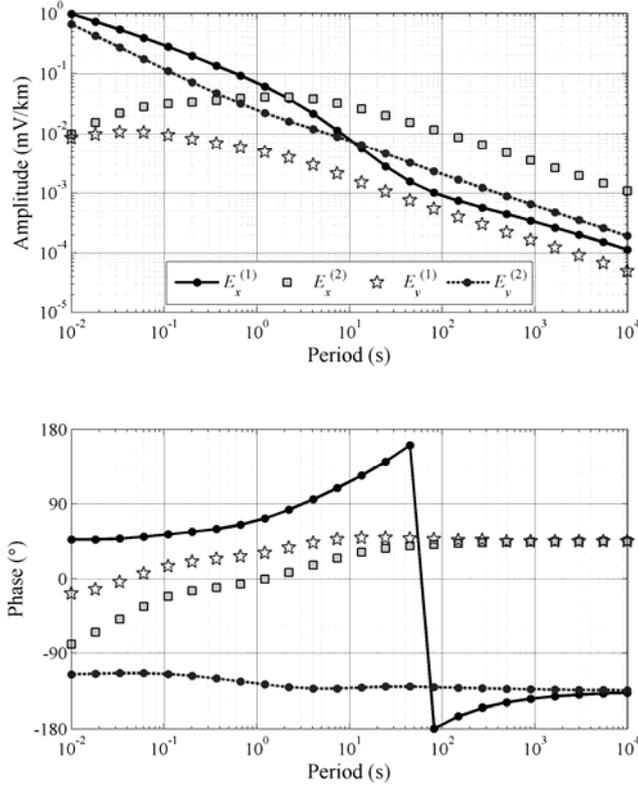

**Figure 7b.** The induced electric field components observed at *Site Y* of the 3-D model shown in Fig. 6. The components $E_x^{(1)}$ and $E_y^{(1)}$ correspond to the fields induced by a unit amplitude magnetic field polarized parallel to the *x*-axis of the model (Polarization 1). The components $E_x^{(2)}$ and $E_y^{(2)}$ correspond to the fields induced by a unit amplitude magnetic field polarized parallel to the *y*-axis (Polarization 2).

Fig. 7c illustrates the characteristic states of the tensor observed at *Site Y*. The top-left panel illustrates the apparent resistivities computed from the eigen-impedances. A noteworthy observation is the significant bay observed in the maximum apparent resistivity curve between 0.1s – 1s, with the overall differential between the highest and lowest values of apparent resistivity approaching 1000Ωm at 0.67s. This is clearly a secondary effect due to the presence of the regional and local conductors and not a structural feature because *site Y* is located at the surface of the homogeneous resistive host. The top-right panel illustrates the phases of the eigen-impedances. It is apparent that the real and imaginary parts of the maximum eigen-impedance are both positive so that the maximum phase is confined in first quadrant. However, *only* for periods $T \notin (1s, 10s)$ are both the real and imaginary parts of the minimum eigen-impedance negative (and the phase contained in the third or first quadrant). For $T \in (1s, 10s)$ the minimum eigen-impedance has a negative part and is *not* passive. The non-passive band coincides with significant variations of the geometry of the electric and magnetic eigen-frames (Fig. 7b, bottom-left and bottom-right). The azimuths $\varphi_E$ and $\varphi_H$ of the electric and magnetic eigen-frames are approx. 0° up to $T \approx 1s$, inasmuch as the "measurements" were conducted parallel and perpendicular to the horizontal axes of geoelectric (intrinsic) coordinate system; at longer periods, $\varphi_E$ declines slightly to the west and $\varphi_H$ rotates to an almost exact east-west orientation (which it attains at $T > 10s$). The tilts $\theta_E$ and $\theta_H$ of the respective eigen-frames are very small up to $T \approx 0.1s$ but explode thereafter and maximize at $T \approx 1.2s$, indicating very steep eigen-field topographies fields and correspondingly high elliptical polarization on the horizontal plane; the magnitude of the tilts diminishes afterwards and returns to normal (horizontal) at long periods, i.e. in the lower homogeneous half-space.

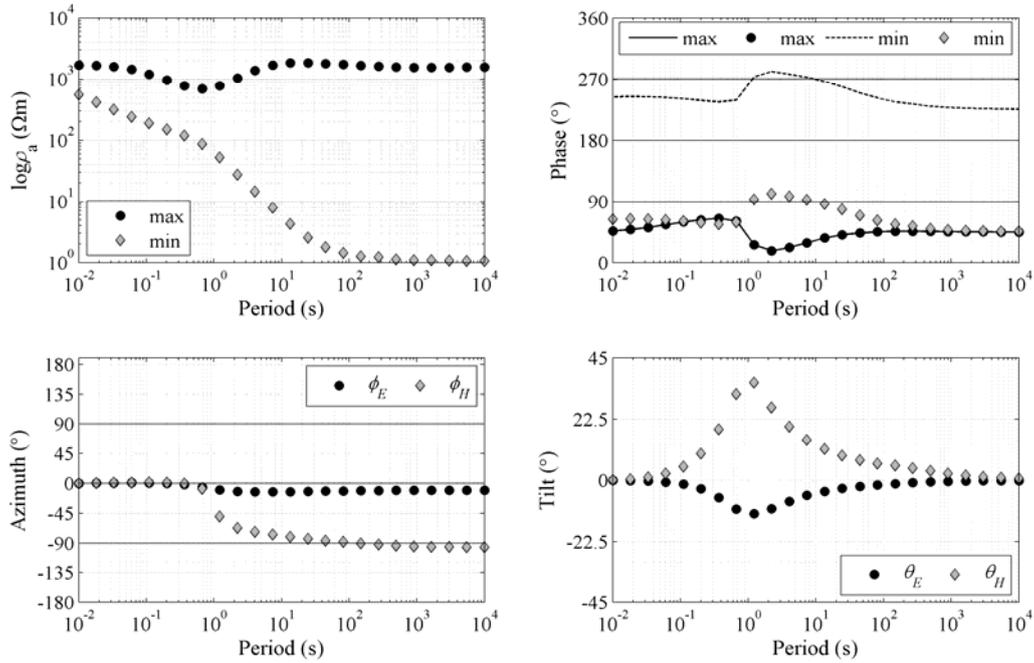

**Figure 7c.** The characteristic states of the impedance tensor observed at *Site Y* of the 3-D model shown in Fig. 6. The layout of the panels and mode of presentation is exactly as per Fig. 2c.

One possibly unexpected observation made in Fig. 7c, was that the breakdown of passivity in the minimum eigen-impedance does not occur at the periods at which it might have been expected by studying the tensor of Fig. 7a, that is between 10s and 1000s, or the eigen-fields (Fig. 7b), and it is not directly associated with the rotation of $E_x^{(1)}$ and $Z_{xy}$ through the second quadrant. Rather, it appears at periods at which $E_x^{(1)}$ and $E_x^{(2)}$ are comparable of strength and interact: specifically it is observed for $T \in$ (1s, 10s), while $E_x^{(2)}$ crosses into the first quadrant, is mostly real and experiences minimal delay, and $E_x^{(1)}$ crosses into the second quadrant, is mostly imaginary (reactive) and experiences maximal delay. This quite unique combination is (expectedly) not limited to the location of *Site Y*, but extends over a relatively significant zone along the western flank of the local conductor (see Figs. 7d – 7f); its possible origin will be examined forthwith.

The investigation will be conducted with hypothetical event analysis based on the eigen-impedances as follows: Assume a real uniform total magnetic eigen-field of the form

$$\tilde{\mathbf{H}}(\theta_H, \varphi_H, \omega) = \begin{bmatrix} H_1(\theta_H, \varphi_H, \omega) \\ H_2\left(\theta_H, \varphi_H + \frac{\pi}{2}, \omega\right) \end{bmatrix} = \begin{bmatrix} 0 & 1 \\ 1 & 0 \end{bmatrix},$$

which induces a hypothetical electric eigen-field

$$\begin{bmatrix} E_1(\theta_E, \varphi_E, \omega) \\ E_2(\theta_E, \varphi_E + \frac{\pi}{2}, \omega) \end{bmatrix} = \begin{bmatrix} 0 & \zeta_1(\omega) \\ -\zeta_2(\omega) & 0 \end{bmatrix} \tilde{\mathbf{H}}(\theta_H, \varphi_H, \omega).$$

Then, counter-rotate the electric eigen-field by an angle $-\varphi_E$ to obtain:

$$\begin{bmatrix} \tilde{E}_x(\theta_E, \omega) \\ \tilde{E}_{y'}(\theta_E, \omega) \end{bmatrix} = \begin{bmatrix} \cos(-\varphi_E) & -\sin(-\varphi_E) \\ \sin(-\varphi_E) & \cos(-\varphi_E) \end{bmatrix} \begin{bmatrix} E_1(\theta_E, \varphi_E, \omega) \\ E_2(\theta_E, \varphi_E + \frac{\pi}{2}, \omega) \end{bmatrix}$$

By this operation, the hypothetical electric eigen-fields are projected on a coordinate frame {$x$, $y'$ $z'$} whose $x$- and $y'$- axes are aligned with the $x$ and $y$ axes of the model (intrinsic coordinate frame) but remains tilted at an angle $\theta_E$ with respect to the intrinsic frame. The hypothetical event analysis will be carried out on the tilted plane by mapping the real and imaginary telluric vectors associated with $\tilde{E}_x(\theta_E, \omega)$ and $\tilde{E}_{y'}(\theta_E, \omega)$, which will henceforth be referred to as the real and imaginary *eigen-telluric vectors* respectively, or RETV/ IETV for short

Fig. 7d shows maps of the RETV (left) and IETV at $T$ = 2.228s, where the highest value of the non-passive minimum eigen-impedance phase is observed in Fig. 7c (the location of *Site Y* is marked with a small open circle). The telluric vectors are superimposed on an outline of the Earth structural model and visualization is further enhanced with streamlines drawn in mid-tone grey colour, which indicate the flow and curvature of the electric eigen-field. In addition, the area in which non-passive minimum phases are observed is marked with light grey shading: it is exclusively located at the west flank of the local conductor and there is no other area with non-passive phases at this period in the entire model.

In the area of interest, to the west of the local conductor, the RETV flows from the regional to the local conductor: the flow starts at a northward direction on the interface of the resistive host with the regional conductor and rotates clockwise until it is almost perpendicular to the western flank of the local conductor, before it abruptly twists to the north at the interface. The maximum curvature of the RETV ro-tation approximately coincides with the $x = 45$km coordinate. In addition, its magnitude increases from S to N. The IETV exhibits analogous behaviour but rotates at a much faster rate and past the $x = 45$km crosses into the SE quadrant before it also twists northwards at the host – local conductor interface. The non-passive minimum phases are observed between $x = 45$km coordinate and the western flank of the local conductor, where they apparently coincide with the zone in which the orientation of the RETV exceeds N45° and the orientation of the IETV exceeds N90° (points to the ESE). This configuration of the RETVs and IETVs, with particular reference to the latter, indicates the existence of high E-W field gradients perpendicular to the local conductor, and, more importantly, current concentration within the resistive host, in an N-S direction parallel to the western flank of the local conductor.

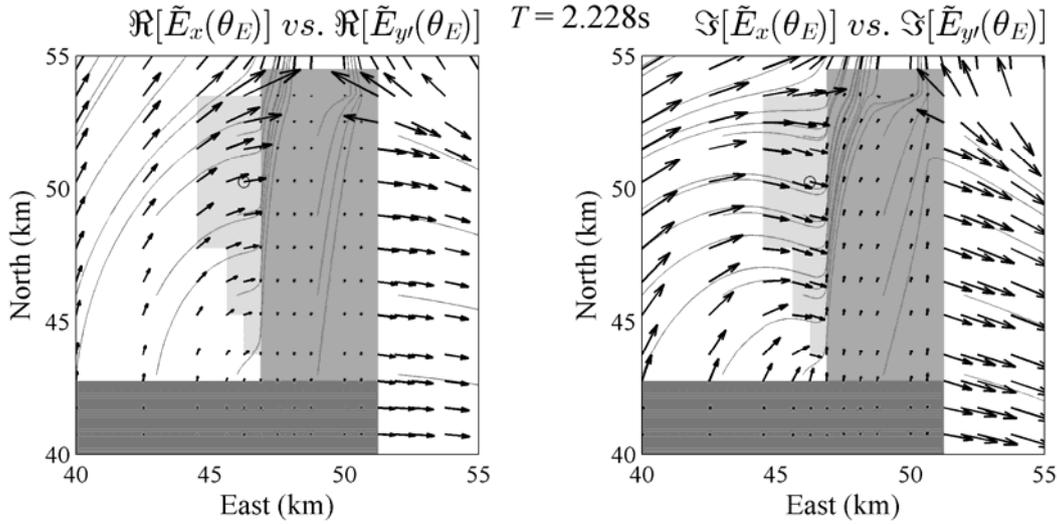

**Figure 7d.** Hypothetical event analysis using the electric eigen-field in the neighbourhood of the local conductor of Fig. 6 and at $T = 2.228$s. The left panel illustrates the telluric vector constructed from the projections of the real parts of the maximum and minimum eigen-fields on the $x$ and $y$ axes of the experimental coordinate frame (*real eigen-telluric vector*). The right panel illustrates the telluric vector constructed from the projections of the imaginary parts of the maximum and minimum eigen-fields on the $x$ and $y$ axes of the experimental coordinate frame (*imaginary eigen-telluric vector*). The visualization flow and curvature of the electric eigen-field is enhanced with mid-tone grey streamlines. The location of *Site Y* is indicated with an open circle.

This observation can be further investigated by studying the gradient of the hypothetical eigen-field, which is shown in Fig. 7e also in the form of eigen-telluric gradient vectors (henceforth ETGV). The top row of Fig. 7e illustrates the *irrotational* part of the eigen-field gradient, with the real part, corresponding to the pair $\Re\{\partial_x \tilde{E}_x(\theta_E)\}$ vs. $\Re\{\partial_{y'} \tilde{E}_{y'}(\theta_E)\}$, shown at the top-left, and the imaginary part, corresponding to the pair $\Im\{\partial_x \tilde{E}_x(\theta_E)\}$ vs. $\Im\{\partial_{y'} \tilde{E}_{y'}(\theta_E)\}$, shown at the top-right. The bottom row of Fig. 7e illustrates the *rotational* part of the eigen-field gradient of which the real part, corresponding to the pair $\Re\{\partial_{y'} \tilde{E}_x(\theta_E)\}$ vs. $\Re\{-\partial_x \tilde{E}_{y'}(\theta_E)\}$ is shown at the bottom-left and the imaginary part, corresponding to the pair $\Im\{\partial_{y'} \tilde{E}_x(\theta_E)\}$ vs. $\Im\{-\partial_x \tilde{E}_{y'}(\theta_E)\}$, is shown at the bottom-right. When drawn this way, the rotational part of the electric eigen-field's gradient is actually a representation of its curl at the direction of the *z′*-axis, i.e. at an angle $\theta_E$ with respect to the *z*-axis of the intrinsic frame.

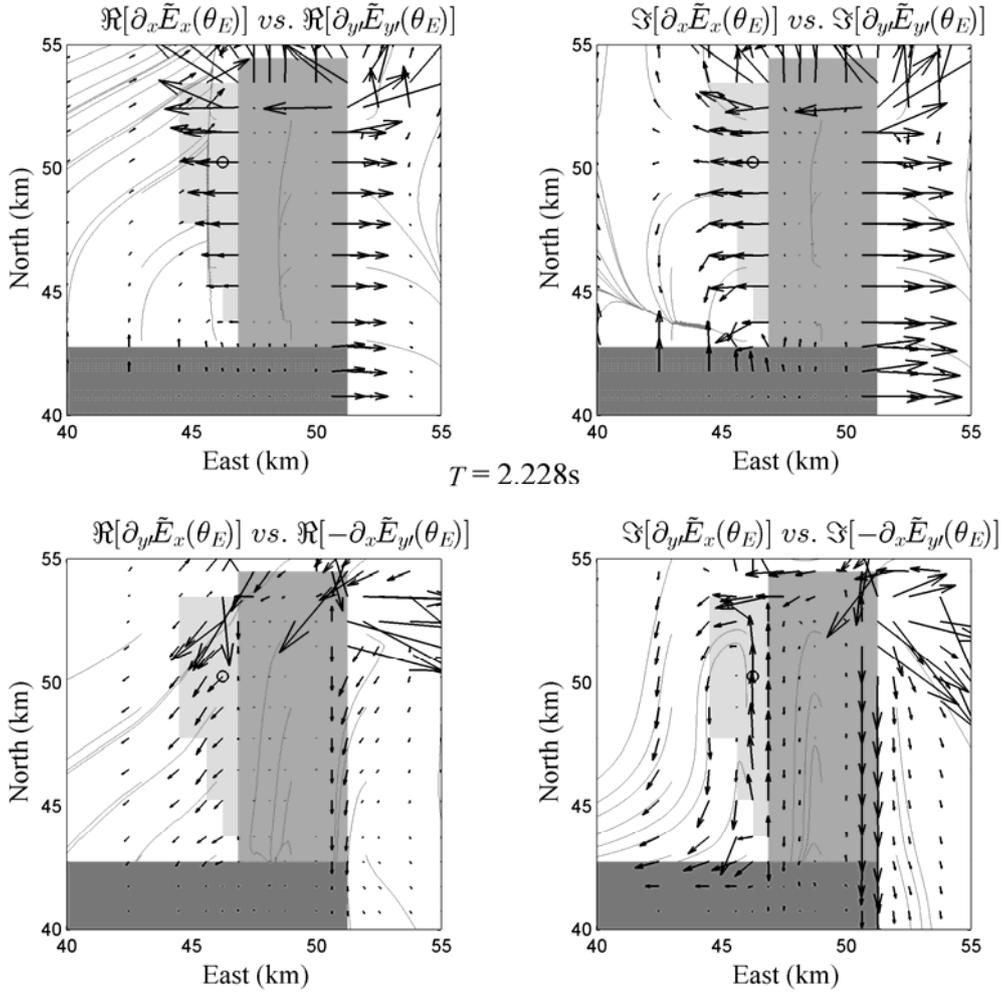

**Figure 7e.** Hypothetical event analysis at $T = 2.228$s using the gradient of electric eigen-field in the neighbourhood of the local conductor of Fig. 6. The top-row panels illustrate the *irrotational* part of the eigen-field gradient, that is $\partial_x \tilde{E}_x(\theta_E)$ vs. $\partial_{y'} \tilde{E}_{y'}(\theta_E)$, as a *real eigen-telluric gradient vector* (top left) and an *imaginary eigen-telluric gradient vector* (top right). The bottom-row panels illustrate the *rotational* part of the eigen-field gradient, that is $\partial_{y'} \tilde{E}_x(\theta_E)$ vs. $-\partial_x \tilde{E}_{y'}(\theta_E)$, also in the form of real (bottom-left) and imaginary (bottom-right) eigen-telluric gradient vectors. Mid-tone grey streamlines are used to enhance visualization of the eigen-field gradient's flow and curvature. The location of *Site Y* is indicated with an open circle.

The configuration of the irrotational component of the ETGV confirms the existence of high gradients normal to the local conductor, whose orientation and magnitude is apparently associated with the zone of non-passive minimum eigen-impedances (Fig 7e, top-left and top-right). Moreover, the real part of the irrotational component indicates a gradient discontinuity just east of the *x*=45km coordinate, clearly observed in the gradient streamlines and also associated with the zone of non-passive minimum eigen-impedances. A more significant observation can be made in the imaginary part of the rotational component of the ETGV, which shows that in the area immediately to the east of the local conductor, there is a N-S oriented counter-clockwise loop of reactive current which is also clearly associated with the zone of non-passive minimum eigen-impedance phases (Fig 7e, bottom-right). Accordingly, a tentative explanation for the violation of passivity involves the combination of high field gradients and the reactive current loop. On one hand, the irrotational part of the gradient may challenge the linearity of the Earth response, although it is static and does not by itself introduce delay. On the other hand, the rotational part is a very suitable agent for the generation of the additional delay required to push the minimum eigen-response into the non-passive realm.

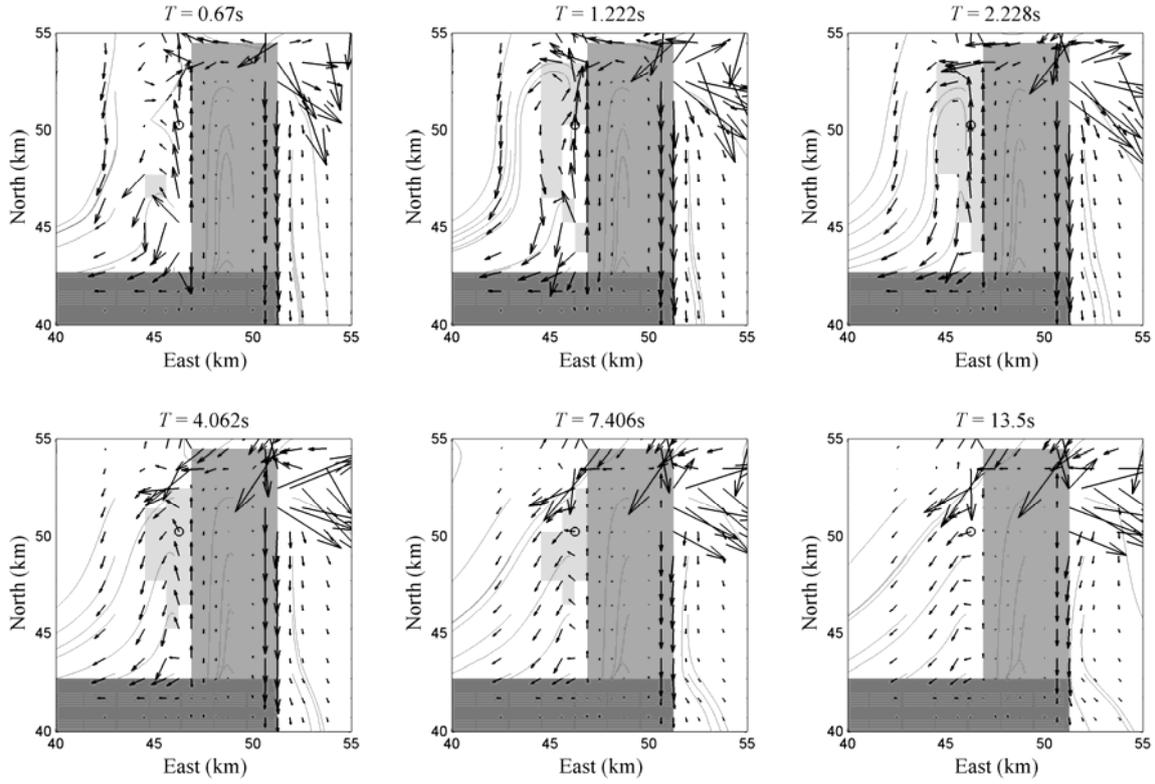

**Figure 7f.** Hypothetical event analysis in the neighbourhood of the local conductor of Fig. 6 and at different periods, using the imaginary rotational eigen-telluric gradient vector, i.e. $\Im\{\partial_{y'}\tilde{E}_x(\theta_E)\}$ vs. $\Im\{-\partial_x\tilde{E}_{y'}(\theta_E)\}$. Mid-tone grey streamlines enhance visualization of the eigen-field gradient's flow and curvature. The location of *Site Y* is indicated with an open circle.

The above explanation may be more convincing if one examines the eigen-field gradient at several periods and compares the conditions under which passivity is violated to the conditions under which it is not. Although a relevant illustration will not be shown for the sake of brevity, it may easily be verified that the configuration of the irrotational part of the gradient does not change significantly with period and does not change in consonance with the area and shape of the zone of non-passive minimum responses. Therefore, the irrotational gradient may comprise a contributing factor, but it certainly is not the decisive one. On the contrary, the imaginary rotational part of the gradient is! This can be clearly seen in Fig. 7f which displays six consecutive instances of the imaginary rotational ETGV, that is $\Im\{\partial_{y'}\tilde{E}_x(\theta_E)\}$ vs. $\Im\{-\partial_x\tilde{E}_{y'}(\theta_E)\}$, in which it may easily be observed that non-passive minimum eig-en-impedance phases are observed always in association with a reactive current loop and only at locations caught between oppositely flowing branches of the loop. Because the tilt of the electric eigen-field changes between locations, the topography of the loop may be variable but it overall forms a cusp with its convex side pointing to the west.

The eigen-impedance analysis of the synthetic data of *Site Y* and adjacent locations documents a case of realizable and realistic linear Earth structure, which upon excitation by a uniform external source field generates a large-scale secondary inductive effect that acts as an internal source and violates the expected passivity of the Earth response. Due to its rather large scale and appearance in an otherwise homogeneous medium, this effect does not qualify as inductive distortion. The non-passive part of the eigen-impedance is apparently not consistent with the tenet of a source-free medium and in principle would comprise a time-invariant non-linear functional of the conductivity structure. Whether this type of response can be interpreted in terms of a realizable model that is consistent with true Earth structure is an altogether different question that will not be addressed herein.

## 5. Summary and Conclusions

The present work attempts to examine some basic and interrelated characteristics and properties of the Magnetotelluric impedance tensor from first principles. The first part of the analysis focuses on the local spatial characteristics of impedance tensors obtained at the surface of general Earth conductivity structures. It is demonstrated that the impedance tensor admits an anti-symmetric generalized eigenstate – eigenvalue decomposition which is consistent with the anti-symmetric interaction between electric and magnetic fields referred to the same coordinate frame. The decomposition is achieved by anti-diagonalization applied through rotation by *two* 2×2 complex operators belonging to the SU(2) rotation group, on the basis of the local isomorphism/ global homeomorphism between rotations in 3-D real space $\mathbb{R}^3$ and rotations in the 2-D Hilbert space $\mathbb{C}^2$, in which SU(2) defines a symmetry group.

The anti-diagonalization involves a left rotation operator whose columns comprise the eigenvectors of the electric field and a right rotation operator whose columns comprise the eigenvectors of the total magnetic field; it yields two characteristic states (generalized eigenstates), with each state comprising a proportional relationship/ interaction between characteristic, non-orthogonal, non-horizontal, linearly polarised input magnetic and output electric field components (the eigen-fields) along the locally fastest (least conductive) and slowest (most conductive) propagation path into the Earth. These interactions are respectively mediated by the maximum and minimum characteristic values of the impedance tensor (the eigen-impedances). The electric and magnetic eigen-fields are the generalized eigenvalues of the transverse electric and magnetic fields observed at the surface of the Earth. The eigen-impedances are simple ratios of electric and magnetic eigen-fields and thus, perfect 3-D generalizations of the scalar impedances familiar from induction in 1-D and 2-D Earth conductivity structures.

The electric and magnetic eigen-fields exhibit 3-D geometry and as noted previously, are generally not orthogonal and not horizontal. This should not come as a surprise because in 3-D Earth structures, the total magnetic and induced electric fields are three dimensional and may be associated with significant gradients, especially in the neighbourhood of structural interfaces: accordingly, they are locally orthogonal and anti-symmetric in 3-space. In this respect, the non-orthogonality of the eigen-fields is only apparent. The local landscape of the electric and total magnetic fields is respectively manifested in the tilt of the electric and magnetic eigen-fields. When the eigen-fields are projected on the axes of the horizontal observational coordinate frame, their components are superimposed: the different mixing of phases effected by the superposition along the orthogonal axes of the experimental coordinate frame manifests itself as elliptical polarization.

The decomposition is closely related to the Singular Value Decomposition of La Torraca et al (1986) and the Canonical Decomposition of Yee and Paulson (1988), which are equivalent and have been constructed on an *ad hoc* basis and from a polarization optics point of view; the relationship can be traced to the fact that the analytical tools of polarization optics, especially those describing spatial transformations, are inherited from the mathematics of spin analysis. However the SVD and CD are *symmetric* decompositions and are defined in a coordinate system in which the orthogonal frames carrying the transverse components of the magnetic and electric fields are not identical but related by a rotation of $\pi/2$. Herein, it was also shown that not only are the SVD and CD proper rotations in $\mathbb{R}^3$, but that they can be mapped onto the coordinate system commonly used in Magnetotelluric data acquisition and analysis, in which the magnetic and electric field frames are identical. The reformulated SVD/CD are indistinguishable from the anti-symmetric decomposition developed herein.

The generalized eigenvalues (eigen-impedances) offer a compact and unique means by which to characterize theoretical and experimental impedance tensors and appraise their suitability for interpretation. Given all the effects and phenomena that may distort Magnetotelluric measurements this would be a particularly helpful utility, but would also require adequate understanding of the conditions under which the eigen-impedances are interpretable and which are generally governed by their analytic structure and properties.

The analytic structure and properties of the eigen-impedances was investigated in Section 3 from first principles, on the basis of uniformity of the external source magnetic field, linearity, pre-Maxwell's equations, energy conservation and the fundamental condition of a source-free linear Earth conductivity structure. No other assumptions were made as to the nature and configuration of the conductivity structure. On the basis of energy conservation it was shown that the eigen-impedances are expected to have all their singularities confined on the positive imaginary frequency and to be positive real functions that obey Kramers – Kronig dispersion relations. In other terms, they are expected to mediate passive induction between freely decaying electric and magnetic fields with positive energy dissipation and their phases to be bounded in the first or third quadrants. As a corollary, the impedance tensor, being the outcome of isometric transformation (linear superposition) of the positive real eigen-impedances, is also expected to be positive real.

It should be emphasized that the positive real property of the eigen-impedances (and measured tensors) is expected, meaning that it is not guaranteed because it is conditional on the absence of sources. Evidently, any location within, or at the surface of the Earth at which *extrinsic* electromagnetic energy is generated and/or injected into the Earth qualifies as a source; this includes all types of natural or anthropogenic noise. A broader definition of a source is given by Chave and Smith (1994) in their analysis of distortion. These authors have shown how distortion is generated by *intrinsic* near-surface secondary effects and that such intrinsic sources may comprise an inductive frequency-dependent (ergo time-dependent) component and a galvanic, frequency-independent component. However, the very definition of passivity, i.e. that the output of a passive physical system at time $t$ must be freely decaying and must depend only on the input at times $\tau \leq t$ clearly indicates that the concept of "source" should be even more general: it should comprise *any* process with energy dissipation characteristics and time dependence sufficiently different than that of free decay and magnitude sufficient as to be able to disrupt it. Conceivably, the scale of inductive processes that may violate passivity can be larger than the scale considered by Chave and Smith (1994), as amply demonstrated in Section 4.3 of the present analysis. An important result of the expanded definition of "internal source" is that passivity violations may be effected by secondary inductive processes in a linear Earth conductivity structure that is totally free of extrinsic effects or small scale inhomogeneities, even when the external forcing magnetic field is uniform. This has an unfortunate repercussion: measured tensor responses that are not positive real are not impedance tensors and cannot be interpreted as such, unless the violating process is identified and/or somehow compensated for.

It is now possible to discuss and possibly clarify the origin of the negative real parts observed in the elements of experimental and synthetic, distorted *or/and* undistorted 3-D impedance tensors. This problem is evidently and intimately related to effects of electric field reversals and the appearance of anomalous phases associated with such reversals.

To begin with, it is important to reassert that time-independent effects cannot violate causality. Galvanic distortion is such an effect – it is practically steady-state and usually described with a real-valued tensor (operator) that twists, shears and differentially amplifies the induced electric field and may even twist it by so much, as to cause reversal of its orientation. Nevertheless, a real operator cannot challenge the passivity of natural (freely decaying) induction because all that it can do is to force a weighted linear superposition of the induced electric field components which, however, remains a weighted linear combination of passive processes and is therefore passive by definition. In other words, real operators do not possess a phase-shifting capacity so as to alter the time-dependence of passive processes. Negative real parts may nevertheless appear in one or more impedance tensor elements because the electric field is a *polar* vector with odd parity, so that an externally forced reversal changes the reference frame of the output electric field, i.e. **E**(**x**, $\omega$) → −**E**(−**x**, $\omega$). A complete reversal introduces a symmetry of $\pi$ in the phase of the electric field and a corresponding shift in the phase of the impedance tensor elements. In consequence, the negative real parts are defined in a reference frame different than that of the measurements and the violation of passivity is only apparent! From a certain point of view, this may is not a particularly welcome outcome because it implies that it would generally not be possible to devise universal analytical methods to discriminate/correct for the effects of galvanic distortion.

Electric field reversals do not only occur as a result of distortion. They may occur as passive effects of natural induction. For passivity to be maintained, the reversal is bound to take place between the first and third quadrants through the fourth, so that the output electric field will always have a positive real part (positive energy dissipation). Such an example is the reversal of the $E_y^{(1)}$ component in Fig. 2b. Reversals may also occur in response to a reactive effect of some internal source (for examples see Sections 4.2 and 4.3). In this case they will take place between the first and third quadrants through the second, in which case the electric field acquires a negative real part (outbound energy flux). However, and in consequence of the principle of superposition and the parallel filter rule, reactively effected reversals must be dominant if they are to have an overall effect on the impedance tensor. It should be re-iterated that in the example of Section 4.3, the breakdown of causality did not have to do with the reversal as such, but only with the parts of the spectrum in which the non-passive reactive effect was dominant or at least very significant in comparison to passive induction modes!

Anomalous phases observed in one or more tensor elements are definitely associated with electric field reversals and in the event of powerful internal sources, with violation of passivity or causality. Nevertheless, due to the dis-predictable combined action of the parallel filter rule and electric reference frame changes during the reversal, the anomalous phases are anything but trustworthy indicators of the true analytic properties of the tensor. To assert whether passivity has been violated or not, it is imperative to refer the output electric field to its *intrinsic* coordinate frame (electric eigen-frame) and evaluate the compliance of the resulting eigen-impedances with the properties expected of an impedance tensor. This can be easily done by utilizing the ASVD and examining the quadrant of the eigen-impedance phases, as shown in Section 4. If the eigen-impedances are passive, then the anomalous phases have only rung a false alarm. Otherwise, the eigen-impedances will turn up with a negative real part, indicating that passivity (and possibly causality) has been violated.

In all cases, the characteristic states and eigen-impedances offer a compact representation of the tensor's constitution and physical properties, as well as unique means of determining what is right or what went wrong. The constraints imposed by the principle of causality on the impedance tensor and the eigen-impedances in particular, comprise fundamental tests for appraising the physical validity of measured data and in this respect, they can be important factors in the process of Magnetotelluric data interpretation.